\documentclass[onecolumn]{aa}

\usepackage[varg]{txfonts}
\usepackage{natbib}
\bibpunct{(}{)}{;}{a}{}{,} 

\usepackage{graphicx} 
\usepackage{dcolumn}
\usepackage{bm}
\usepackage{amssymb}
\usepackage{amsmath}
\usepackage{wasysym}
\usepackage{enumerate}
\usepackage{color}
\usepackage{hyperref}
\usepackage{epstopdf}

\usepackage{float}
\usepackage{flushend}

\usepackage{hyperref}
\hypersetup{
    colorlinks,%
    citecolor=blue,%
    filecolor=blue,%
    linkcolor=blue,%
    urlcolor=blue
}

\begin{document}


\title{A 3D kinematic Babcock Leighton solar dynamo model sustained by dynamic magnetic buoyancy and flux transport processes}


\author{Rohit Kumar \inst{1,2} \thanks{This work is dedicated to the memory of Rohit Kumar, the first author of this paper.} \and Laur{\`e}ne Jouve \inst{1} \and Dibyendu Nandy \inst{2,3}}

 \institute{Institut de Recherche en Astrophysique et Plan{\'e}tologie, Universit{\'e} de Toulouse, CNRS, UPS, CNES, 31400 Toulouse, France \\
           \email{\bf{ ljouve@irap.omp.eu}}
           \and 
          Center of Excellence in Space Sciences India, Indian Institute of Science Education and Research Kolkata, Mohanpur 741246, West Bengal, India
          \email{dnandi@iiserkol.ac.in}
          \and 
          Department of Physical Sciences, Indian Institute of Science Education and Research Kolkata, Mohanpur 741246, West Bengal, India
          } 
          

\abstract
{Magnetohydrodynamic interactions between plasma flows and magnetic fields is fundamental to the origin and sustenance of the 11-year sunspot cycle. These processes are intrinsically three-dimensional (3D) in nature.}
{Our goal is to construct a 3D solar dynamo model that on the one hand captures the buoyant emergence of tilted bipolar sunspot pairs, and on the other hand produces cyclic large-scale field reversals mediated via surface flux-transport processes -- that is, the Babcock-Leighton mechanism. Furthermore, we seek to explore the relative roles of flux transport by buoyancy, advection by meridional circulation, and turbulent diffusion in this 3D dynamo model.}
{We perform kinematic dynamo simulations where the prescribed velocity field is a combination of solar-like differential rotation and meridional circulation, along with a parametrized turbulent diffusivity. We use a novel methodology for modeling magnetic buoyancy through field-strength-dependent 3D helical up-flows that results in the formation of tilted bipolar sunspots.}
{The bipolar spots produced in our simulations participate in the process of poloidal-field generation through the Babcock-Leighton mechanism, resulting in self-sustained and periodic large-scale magnetic field reversal. Our parameter space study varying the amplitude of the meridional flow, the convection zone diffusivity, and parameters governing the efficiency of the magnetic buoyancy mechanism reveal their relative roles in determining properties of the sunspot cycle such as amplitude, period, and dynamical memory relevant to solar cycle prediction. We also derive a new dynamo number for the Babcock-Leighton solar dynamo mechanism which reasonably captures our model dynamics.}
{This study elucidates the relative roles of different flux-transport processes in the Sun's convection zone in determining the properties and physics of the sunspot cycle and could potentially lead to realistic, data-driven 3D dynamo models for solar-activity predictions and exploration of stellar magnetism and starspot formation in other stars.}


\titlerunning{A 3D Babcock Leighton Solar Dynamo Model}

\authorrunning{Kumar, Jouve and Nandy}

\maketitle

\section{Introduction}
\label{intro}

The solar magnetic cycle and the magnetized sunspots that it generates eventually determine many aspects of solar activity such as flares and coronal mass ejections, solar irradiance, solar wind, and heliospheric open flux variations ~\citep{Ossendrijver:AAR2003}. These physical processes are major contributors to what is now commonly called space weather and can significantly impact our space environment and planetary atmospheres. Apart from long-term observations obtained both through ground- and space-based telescopes which constrain the spatio-temporal evolution of solar surface magnetic fields, theoretical insights and numerical simulations based on magnetohydrodynamic (MHD) dynamo models have proven to be very effective in understanding the origin and evolution of magnetic fields in the interior of the Sun.

The toroidal component of magnetic fields that generate sunspots are understood to be produced by stretching of poloidal field lines by the differential rotation of the Sun \citep{Parker:APJ1955b}. Strong toroidal flux tubes become magnetically buoyant \citep{Parker:APJ1955a} and rise through the solar convection zone producing bipolar pairs which are tilted due to the action of the Coriolis force on the rising flux tubes \citep{D'Silva:AA1993}. Although a variety of mechanisms have been proposed for regeneration of the poloidal component of the Sun's magnetic field \citep{Charbonneau:LRSP2005}, observational analysis of solar-cycle properties support the Babcock-Leighton {\bf (BL)} mechanism as the dominant poloidal field source \citep{Dasi-Espuig:AA2010,MunozJaramillo2013}. The Babcock-Leighton mechanism \citep{Babcock:APJ1961,Leighton:APJ1969} involves the action of near-surface flux-transport mechanisms on buoyantly emerged tilted bipolar sunspot pairs, which results in cross-equatorial flux cancelation of leading polarity sunspots and migration of the following polarity flux towards the poles to reverse the large-scale solar polar field (associated with the poloidal component). The newly generated poloidal field is then subducted by a variety of flux transport processes and subject to shearing by differential rotation which produces the toroidal field of the subsequent sunspot cycle. These models are amenable to solar surface data assimilation and can be used in predictive modes to make predictions of future solar activity. 

Solar dynamo modeling is carried out by adopting mainly two types of approaches: two-dimensional (2D) kinematic mean-field models \citep{Moffatt:book,Radler:AA1990} in which for a prescribed velocity field, the evolution of the large-scale magnetic field is studied \citep{Dikpati:APJ1999,Nandy:APJ2001,Jouve:AA2007} and three-dimensional (3D) global MHD models where the full set of MHD equations is self-consistently solved \citep{Nelson:APJ2013,Miesch:ARFM2009,Brun:SSR2015}. In the 2D mean-field kinematic models, a Babcock-Leighton source term can be added to the poloidal field evolution equation to model the rise of flux tubes from the base of the convection zone to the solar surface where sunspot pairs are produced. However, in these axisymmetric models, the rise of flux tubes is treated as instantaneous. Various levels of treatment have evolved in sophistication, starting from the direct addition of a net effective azimuthally averaged poloidal field close to the near-surface layer to placing double-rings in terms of the axisymmetric vector potential for the poloidal field at near-surface layers to mimic the eruption of bipolar sunspot pairs \citep{Durney1995,Durney1997,Nandy:APJ2001,MunozJaramilloetal2010}. Although these models reproduce large-scale solar-cycle features reasonably well, they do not adequately capture the actual production of non-axisymmetric tilted bipolar magnetic regions (BMRs) or how they result in this effective poloidal flux -- processes that are intrinsically 3D.  We note that new models have been developed recently coupling a 2D axisymmetric dynamo model to a 2D surface flux-transport model~\citep{Lemerle:APJ2015,Lemerle:APJ2017}. This new idea now indeed enables the production of BMRs from the toroidal field of the dynamo simulation and allows for their evolution through the surface flux-transport code to be followed. In 3D global MHD models, naturally buoyant twisted magnetic structures do start to be self-consistently produced in the convection zone~\citep{Nelson:APJ2013,Fan:APJ2014}, but their density deficit is still erased too quickly to maintain coherent structures rising all the way to the top of the domain to create active regions. These models do not produce systematically tilted BMRs satisfying the observed Joy's law distribution of tilt angles. Moreover, 3D global MHD models are computationally expensive for these types of studies and do not easily allow for a parametric study to be performed. Therefore there is a need to further develop these approaches to overcome some of these deficiencies. 

One crucial element in this context is a relatively more sophisticated modeling of the buoyant emergence of tilted flux tubes within the framework of the Babcock-Leighton mechanism. Recently, kinematic 3D flux-transport models have been developed \citep{Miesch:APJL2014,Yeates:MNRAS2013}. These models are obviously more realistic than 2D axisymmetric ones since non-axisymmetric tilted bipolar magnetic regions are created and can actively participate in the production of a net poloidal magnetic flux at the surface. The resultant surface magnetic flux gets advected by the surface flows and transported by diffusion towards the poles to reverse their polarities. \cite{Cameron:Science2015} have demonstrated analytically the connection between the generation of the toroidal flux in the two hemispheres and the shearing and amplification of the surface poloidal flux. In their 3D dynamo models, \cite{Miesch:APJL2014} and \cite{Miesch:ASR2016} adapted a double-ring algorithm in which BMRs are directly placed at the solar surface. In these types of models however the flux emergence in the convection zone is missing, which is believed to be an important aspect of the solar magnetic cycle. On the other hand, \cite{Yeates:MNRAS2013} adopted a different approach where they employed an additional radial velocity in the convection zone that transports the toroidal flux from the base of the convection zone to the surface to produce bipolar spots. However, they focussed only on one magnetic cycle, and did not demonstrate the ability of their model to achieve sustained long-term cyclic dynamo action. In a new 3D flux-transport solar dynamo model, \cite{Kumar:Frontiers2018} implemented a magnetic buoyancy algorithm similar to that by \cite{Yeates:MNRAS2013} with the potential of sustaining long-term dynamo action.  In this paper, we develop and refine this model with 3D helical flows to model the flux emergence process and show that this model produces self-sustained dynamo action fed by flux-transport processes such as magnetic buoyancy, meridional circulation, and turbulent diffusion.  We perform an extended parametric study of this model to assess the dependence of magnetic cycle properties on the parameters governing the efficiency of various flux transport processes. 
  
Long-term sunspot observations  ~\citep{Hathaway:LRSP2010} show variations in the duration and amplitude of the solar cycle, including grand minima episodes whose origin may be in fluctuations of the dynamo source term \citep{Passosetal2014,Hazra:APJ2014} or flux transport parameters. While some studies infer that meridional circulation variations may play an important role in this \citep{Charbonneau:APJ2000,Hathaway:APJ2003}, other studies indicate that turbulent pumping may also sustain similar cycle dynamics \citep{HazraNandy2016}. The propagation of these fluctuations to the next cycles through flux-transport processes is an emergent issue with critical relevance for solar-cycle predictions \citep{Yeates:APJ2008,KarakNandy2012}. In an attempt to anticipate the strength of cycle 24, \cite{Dikpati:APJ2006} used an advection-dominated model (strong meridional circulation and low turbulent diffusion) and showed that the toroidal field of cycle $n$ is produced by the combined poloidal fields of cycles $n-3, n-2$, and $n-1$ and predicted that cycle $24$ would be a very strong cycle~\citep{Dikpati:GRL2006}. On the other hand, \cite{Choudhuri:PRL2007} used a diffusion-dominated model and showed that the toroidal field of cycle $n$ is generated mainly by the poloidal field of cycle $n-1$ and that cycle $24$ would be a very weak cycle. \cite{Yeates:APJ2008} explained the differing dynamical memory of these models and the consequent diverging predictions on the basis of the relative roles of flux-transport by meridional circulation and turbulent diffusion and they demonstrated in their 2D dynamo model that the dynamo behaves very differently in this regime.  Inspired by these studies, we perform a detailed parameter-space study to understand magnetic fields dynamics in our 3D Babcock-Leighton dynamo model. Specifically, we vary the amplitude of the meridional circulation, the convection zone diffusivity, and, in addition, parameters related to BMR emergence, and study how these variations affect the magnetic cycle. We also present a new dynamo number which may be suitable for characterizing dynamo action in Babcock-Leighton dynamo models of the solar cycle. 

The paper is organized as follows: In Sect.~\ref{sec:fl_tr_mdel}, we present the 3D flux-transport kinematic dynamo model. Results of dynamo simulations and the magnetic cycle are shown in Sect.~\ref{sec:dyn_sol_mag_cyc}. In Sect.~\ref{sec:para_study}, effects of various parameters on the magnetic cycle are discussed. Finally, in Sect.~\ref{sec:conclusion}, we summarize our results.

\section{The 3D flux-transport dynamo model}
\label{sec:fl_tr_mdel} 

\subsection{The numerical framework}
\label{sec:num_detail} 

We solve the magnetic induction equation for our kinematic dynamo simulation, which is as follows~\citep{Davidson:book}. 
\begin{eqnarray}
\frac{\partial \mathbf{B}}{\partial t} = \nabla \times (\mathbf{v} \times \mathbf{B}) - \nabla \times (\eta \nabla \times \mathbf{B}),
\end{eqnarray}
where $\mathbf{B}$ is the magnetic field, $\mathbf{v}$ is the prescribed velocity field, and $\eta$ is the magnetic diffusivity. In our simulations, the prescribed velocity field is a combined effect of the solar-like differential rotation and the meridional circulation, and a velocity corresponding to the magnetic buoyancy in the convection zone. Here, the velocity field is defined as
\begin{eqnarray}
\mathbf{V} = V_r \overrightarrow{e_r} + V_\theta \overrightarrow{e_\theta} + r \sin\theta\, \Omega \overrightarrow{e_\phi} + \overrightarrow{V_b}, 
\end{eqnarray} 
where $(V_r, V_\theta)$ is the axisymmetric meridional flow, $\Omega$ is the spatially dependent rotation rate, and $\overrightarrow{V_b}$ is the buoyancy velocity. Each of these velocity field components is described below.

We solve the magnetic induction equation in a spherical shell using the publicly available pseudo-spectral solver MagIC~\citep{Wicht:PEPI2002,Gastine:Icarus2012} which can de downloaded on \verb?https://github.com/magic-sph?. MagIC uses a poloidal/toroidal decomposition both for the magnetic field and the mass flux to ensure the divergence constraint. MagIC employs spherical harmonic decomposition in the azimuthal and latitudinal directions, and Chebyshev polynomials in the radial direction. For time-stepping, it employs semi-implicit Crank-Nicolson scheme for the linear terms and Adams-Bashforth for the nonlinear terms. The inner and outer radii of computational shell are $[0.65, 1.0] \, R_{\odot}$, where $R_{\odot}$ is the solar radius. We choose $N_r=64$ grid-points in the radial, $N_\theta=128$ points in the latitudinal, and $N_\phi=256$ points in the longitudinal directions. Simulations are performed by considering insulating boundaries of the spherical shell. As an initial condition, we employ a dipolar magnetic field.

\subsection{Axisymmetric velocity field}   
\label{sec:vel_field}

\begin{figure*}[htbp]
\centering
\includegraphics[scale=1.1]{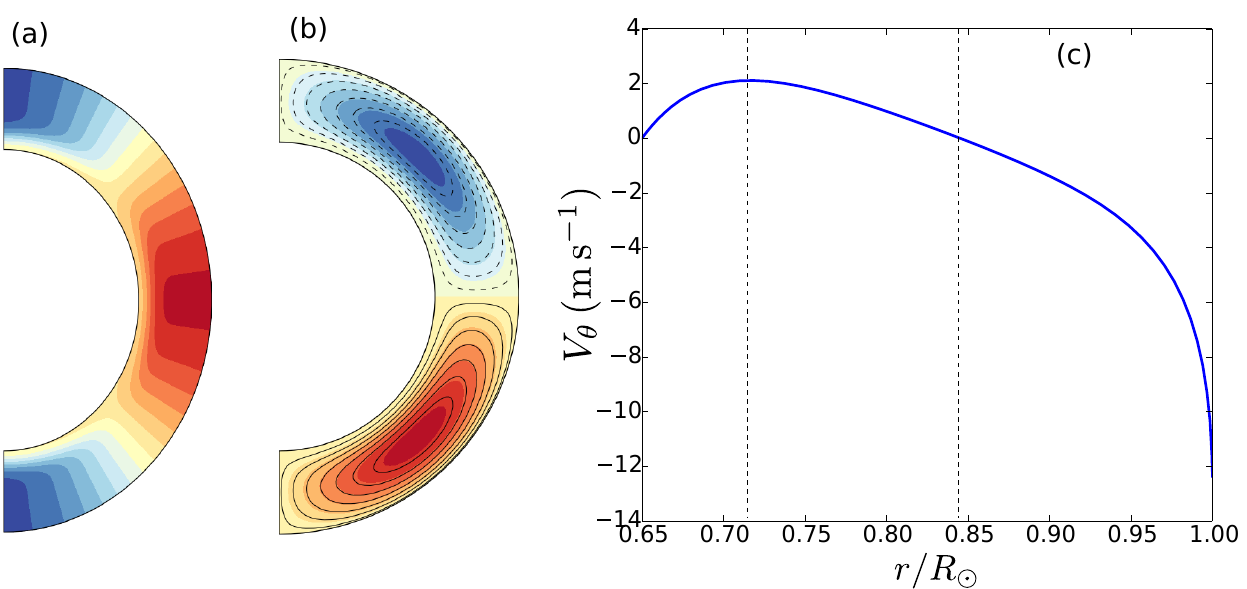}
\caption{The plots of axisymmetric velocities: (a) differential rotation and (b) meridional circulation (dashed lines: anti-clockwise, solid lines: clockwise), and (c) the latitudinal velocity ($V_\theta$) with normalized radius (at $\theta=45^\circ$) in the convection zone.}
\label{fig:diff_mer_fl_num}
\end{figure*}

We prescribe an axisymmetric velocity profile which is a combination of a differential rotation and a meridional circulation. The rotation profile is similar to that observed in the Sun through helioseismology~\citep{Schou:APJ1998}. The normalized rotation rate is defined as
\begin{eqnarray}
\Omega(r,\theta) &=& \Omega_c + \frac{1}{2} \left[ 1 + erf \left(\frac{r - r_c}{d_1} \right)  \right]  \left( 1 - \Omega_c - c_2 \, \cos^2 \theta\right),
\end{eqnarray}
where $\Omega_c= 0.92, c_2=0.2, r_c = 0.7 \, R_\odot$ (the tachocline), and $d_1=0.15 \, R_\odot$. The rotation is strongest in the equatorial region, and it decreases as we move towards the poles [see Fig.~\ref{fig:diff_mer_fl_num}(a)]. A similar rotation profile was used by \cite{Jouve:AA2008}.

 In addition, the flow also consists of a single-cell meridional circulation (MC) per hemisphere, which is poleward near the outer surface and equatorward near the base of the convection zone [shown in Fig.~\ref{fig:diff_mer_fl_num}(b)]. The meridional circulation plays an important role in BL flux-transport models where it advects the effective magnetic flux of the trailing spots towards the poles to reverse the polarities of the polar field~\citep{Wang:APJ1989,Dikpati:AA1994,Dikpati:SP1995,Choudhuri:SP1999}. We use a meridional circulation profile similar to that of \cite{Dikpati:APJ1999} with a deeply penetrating component of the flow to just beneath the base of the convection zone as suggested by \cite{Nandy:SCIENCE2002}. Such a flow achieves adequate coupling of various layers of the convection zone and ensures low latitude activity which is otherwise very difficult to achieve. For the normalized density profile in the convection zone defined as $\rho(r)= \left[(R_\odot / r) - 0.95 \right]^m$, the normalized radial and latitudinal components of the meridional flow are as follows 
\begin{eqnarray}
V_r(r,\theta) &=& \left(\frac{R_\odot}{r}\right)^2 \left[ - \frac{1}{m+1} + \frac{l_1}{2m+1} \, \xi^m - \frac{l_2}{2m+p+1} \, \xi^{m+p}  \right] \nonumber \\ 
                   \; &\times& \; \xi \; (2 \cos^2\theta - \sin^2\theta),  \\
V_\theta(r,\theta) &=& \left(\frac{R_\odot}{r}\right)^3 \left[ - 1 + l_1 \, \xi^m - l_2 \, \xi^{m+p}  \right] \, \sin \theta \; \cos \theta,
\end{eqnarray}
where 
\begin{eqnarray}
l_1 &=&  \frac{(2m+1)(m+p)}{(m+1)p} \xi_b^{-m}, \\
l_2 &=&  \frac{(2m+p+1)m}{(m+1)p} \xi_b^{-(m+p)}, \\
\xi(r) &=& \frac{R_\odot}{r} -1,
\end{eqnarray}
$\xi_b=0.54$, $m=0.5$, and $p=0.25$. \\

The observations suggest that the magnitude of the rotational velocity at the equator is around $2 \; \rm km \, \rm s^{-1}$ and the peak meridional flow at the solar surface is approximately $20 \; \rm m \, \rm s^{-1}$~\citep{Roudier:AA2012}. Considering these values, we fix the maximum longitudinal velocity to $2 \; \rm km \, \rm s^{-1}$ (at the surface at the equator) and vary the peak surface latitudinal velocity in the range $7.5 - 20 \; \rm m \, \rm s^{-1}$ in different cases since the MC is a much more fluctuating flow and we want to assess the effects of a varying the MC amplitude on the magnetic field evolution. The variation of latitudinal velocity with radius is illustrated in Fig.~\ref{fig:diff_mer_fl_num}(c), where the maximum surface velocity is $12.5 \; \rm m \, \rm s^{-1}$  at latitudes $\theta = \pm 45^\circ$.   

\subsection{Magnetic diffusivity}  
\label{sec:mag_diff}

\begin{figure}[htbp]
\centering
\includegraphics[scale=0.4]{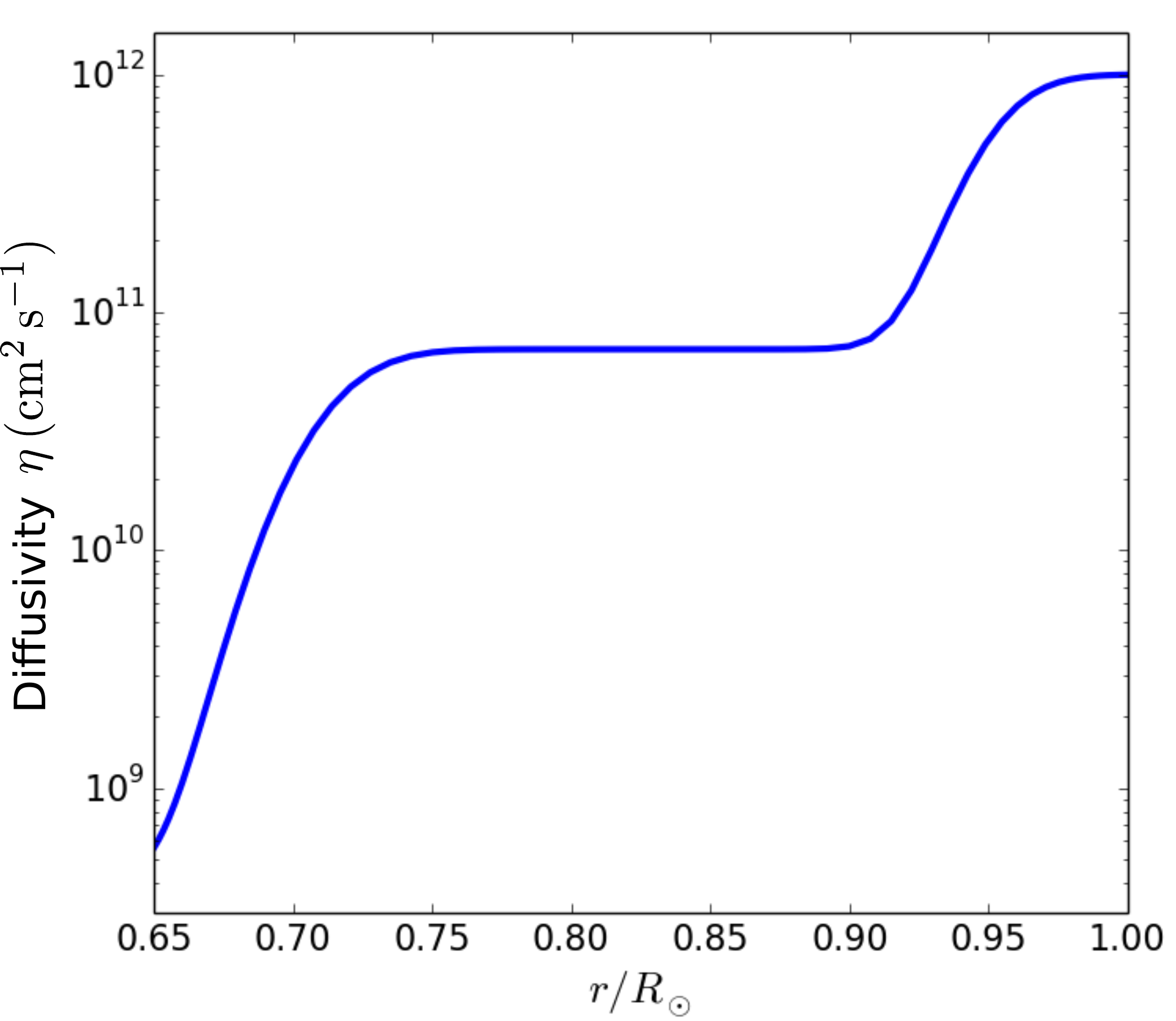}
\caption{The two-step magnetic diffusivity as a function of normalized radius in the convection zone. Here the core diffusivity $\eta_0= 4 \times 10^{8} \, \rm cm^2 \, \rm s^{-1}$, the convection zone diffusivity $\eta_c= 7 \times 10^{10} \, \rm cm^2 \, \rm s^{-1}$, and the surface diffusivity $\eta_s= 10^{12} \, \rm cm^2 \, \rm s^{-1}$.}
\label{fig:mag_diff}
\end{figure}

The magnetic diffusivity is chosen as a two-step function such that the diffusivity near the inner core is small, larger in the convection zone, and maximum near the outer surface, similar to that of \cite{Yeates:MNRAS2013}. The magnetic diffusivity is defined as
\begin{eqnarray}
\eta(r) = \eta_0 + (\eta_c-\eta_0) \left[ 1 + {\rm erf} \left(\frac{r - R_5}{d_6} \right) \right] + (\eta_s - \eta_c-\eta_0) \left[ 1 + {\rm erf} \left(\frac{r - R_6}{d_7} \right)\right] ,\label{eq:eta}
\end{eqnarray}
where the inner core diffusivity $\eta_0= 4 \times 10^{8} \, \rm cm^2 \, \rm s^{-1}$, the surface diffusivity $\eta_s= 10^{12} \, \rm cm^2 \, \rm s^{-1}$, and the convection zone diffusivity $\eta_c$ is varied in different simulations. The other parameters are fixed to $r_5 = 0.71 \,R_\odot, r_6=0.95 \,R_\odot, d_6=0.03 \,R_\odot$, and $d_7=0.025 \,R_\odot$. The profile of the magnetic diffusivity with radius in the convection zone is illustrated in Fig.~\ref{fig:mag_diff}, where $\eta_c= 7 \times 10^{10} \, \rm cm^2 \, \rm s^{-1}$. 

Having described the prescribed axisymmetric velocity field, the initial conditions, and the diffusivity profile for the simulations, we now discuss the buoyancy algorithm in the convection zone.

\subsection{Magnetic buoyancy algorithm}
\label{sec:buoy_algo}

In this section, we present the magnetic buoyancy algorithm implemented in our model which is crucial for BMR generation at the surface. For this purpose, we employ a buoyancy velocity field in the convection zone that acts in the radially outward direction. The buoyancy velocity transports the toroidal flux from the bottom of the convection zone to the outer surface. The buoyancy velocity as a function of longitude and latitude (${\phi}, {\theta}$) is described as in~\citep{Yeates:MNRAS2013} and in our previous work \citep{Kumar:Frontiers2018} by
\begin{equation}
V_b = V_{b0} \; exp{\left[-\left\lbrace \left(\frac{\phi - {\bar \phi}}{\sigma_\phi}\right)^2 + \left(\frac{\theta - {\bar \theta}}{\sigma_\theta}\right)^2 \right\rbrace \right]},
\end{equation}   
where (${\bar \phi}, {\bar \theta}$) corresponds to the apex of the rising flux tubes, which moves with the local rotation rate, and the extension of the buoyant parts of the flux tubes are chosen to be $\sigma_\theta = \sigma_\phi =5$ degrees. This is motivated by studies of magnetic buoyancy instabilities of flux sheets which produce tube-like structures of relatively small scale when triggered~\citep{Parker:APJ1955b,Cattaneo:JFM1988,Matthews:APJ1995,Jouve:APJ2013}. The amplitude of the buoyancy velocity, $V_{b0}$, is uniform in all runs except the one discussed in Sect.~\ref{sec:tor_buoy} and is set to $94.5 \, \rm m \, \rm s^{-1}$ such that it takes about one month to transport the toroidal field from the bottom of the convection zone to the surface. In our model, the values of (${\bar \phi}, {\bar \theta}$) are chosen randomly so that ${\bar \phi}$ can take any value in the range $[0^\circ- 360^\circ]$ longitude, but ${\bar \theta}$ lies in the latitudinal region $[-35^\circ, +35^\circ]$, the latitudes of observed bipolar spots at the solar surface~\citep{Maunder:MNRAS1922}. As an effect of the buoyancy velocity, the toroidal flux ropes at the bottom of the convection zone start to emerge in the radially outward direction to produce the bipolar magnetic regions at the surface. The additional buoyancy velocity field acting at a particular (${\bar \phi}, {\bar \theta}$) is applicable only if the toroidal magnetic field $B_\phi > B_{\phi}^l$ ($=4 \times 10^4$ gauss), which is in agreement with the physics of magnetic buoyancy instabilities. Indeed, it is thought that below this value, the field is either too small to be affected by the magnetic buoyancy or if buoyant enough, strongly influenced by the Coriolis force, and hence rises parallel to the rotation axis~\citep{Choudhuri:APJ1987}. As we are working with a kinematic dynamo model, we suppress the effects of the magnetic buoyancy for $B_\phi > B_{\phi}^h$ ($=1.4 \times 10^5$ gauss) in order to achieve energy saturation. This is physically justified by the fact that stronger fields are expected to rise briskly to the surface without being affected by the Coriolis force and as a result, the produced BMRs would not be tilted enough and hence would not participate in the polar field reversals. To avoid additional complexities we do not include any radial dependence that precludes flux-tube expansion. This is done with the understanding that it is the total flux content rather than field strength of emerged bipoles that is relevant for the BL mechanism.

\begin{figure}[htbp]
\centering
\includegraphics[scale=0.5]{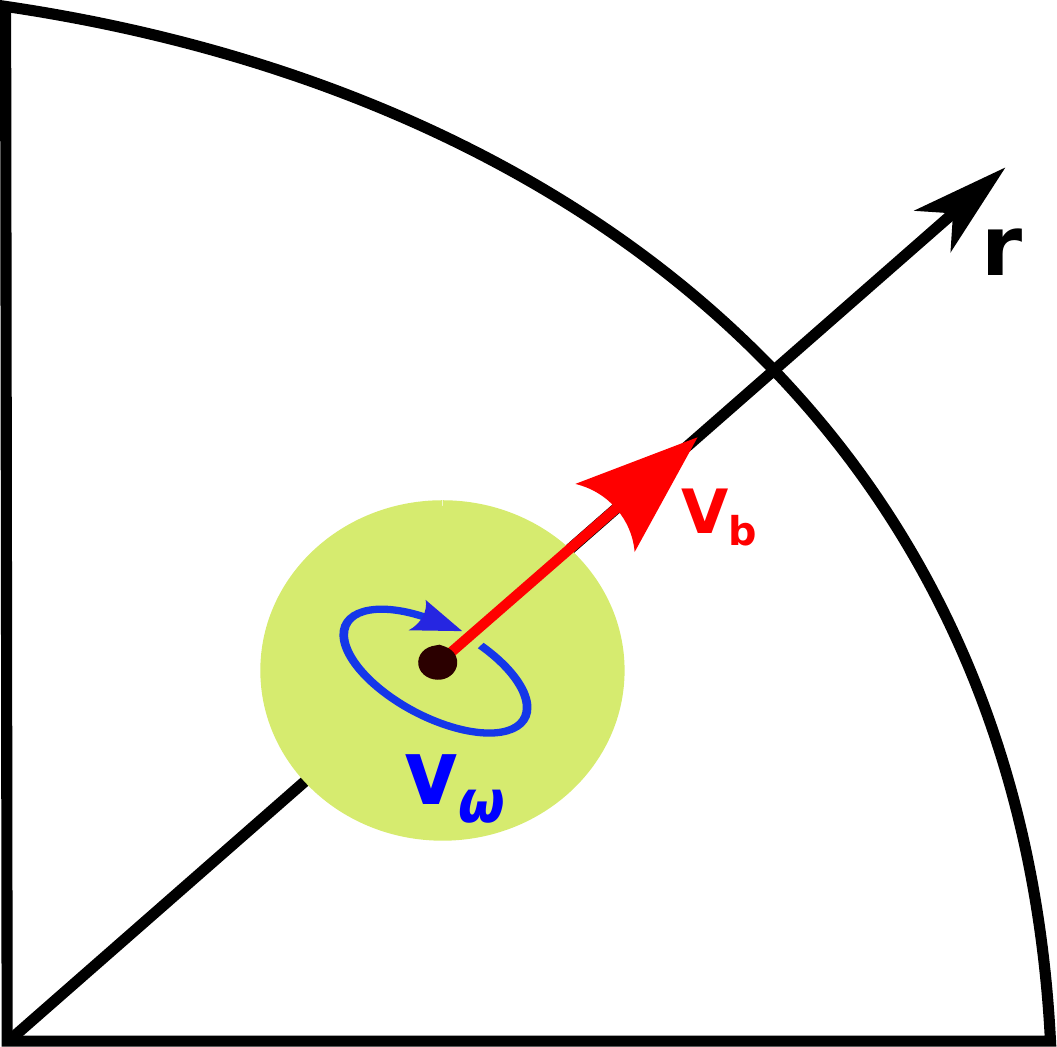}
\caption{Schematic diagram for the radial velocity corresponding to buoyancy ($V_b$) and vortical velocity ($V_\omega$).}
\label{fig:vr_omega}
\end{figure}

To include the effect of the Coriolis force on toroidal fields between $B_{\phi}^l$ and $B_{\phi}^h$, we then apply an additional vortical velocity, which imparts spatial twist in the emerging magnetic flux ropes. The vortical velocity ($V_\omega$) is the combination of a latitudinal and a longitudinal velocity component, defined as ~\citep{Yeates:MNRAS2013}
\begin{eqnarray}
V^t_{\theta} &=& \frac{V^t_{\theta0}}{2} \; r \cos{\bar \theta} \; \sin{\bar \theta} \; e^{- \zeta^2/\delta^2} \; \sin(\phi - {\bar \phi}), \label{eq:v_th} \\ 
V^t_{\phi} &=& - \frac{V^t_{\phi0}}{2} \; r \cos{\bar \theta} \; e^{- \zeta^2/\delta^2} \nonumber \\
&\times&\; [\sin\theta \; \cos{\bar \theta} 
 - \cos\theta \; \sin{\bar \theta} \; \cos(\phi - {\bar \phi})],\label{eq:v_ph} 
\end{eqnarray}
where $V^t_{\theta0}$ and $V^t_{\phi0}$ are the amplitudes of the latitudinal and the longitudinal velocities, respectively,
\begin{equation}
\zeta = \sqrt{r^2 + {\bar r}^2 -2r{\bar r} (\sin\theta \, \sin{\bar \theta} \, \cos(\phi - {\bar \phi}) + \cos\theta \cos{\bar \theta})} 
,\end{equation}
which is the Euclidean distance of a spatial point ($r, \theta, \phi$) from the center (${\bar r}, {\bar \theta}, {\bar \phi}$), \\
$\delta = \delta_0 \sqrt{\frac{R_{\odot}/{\bar r}_0 - 0.95}{R_{\odot}/{\bar r} - 0.95}}$, and $\delta_0=(5\pi/18)(0.7R_{\odot})$. Here $\delta$ represents the radius of the velocity components $V^t_{\theta}$ and $V^t_{\phi}$ , and $\delta_0$ is the initial radius at ${\bar r} = {\bar r}_0$. In Fig.~\ref{fig:vr_omega}, we illustrate a schematic diagram for radial and vortical velocities: $V_b$, $V_\omega(V^t_{\theta},V^t_{\phi})$.

The combination of these radial and vortical velocity fields effectively yields a helical flow which mimics or models the buoyant emergence of tilted toroidal flux tubes from the solar interior. 

\begin{figure}[htbp]
\centering
\includegraphics[scale=0.3]{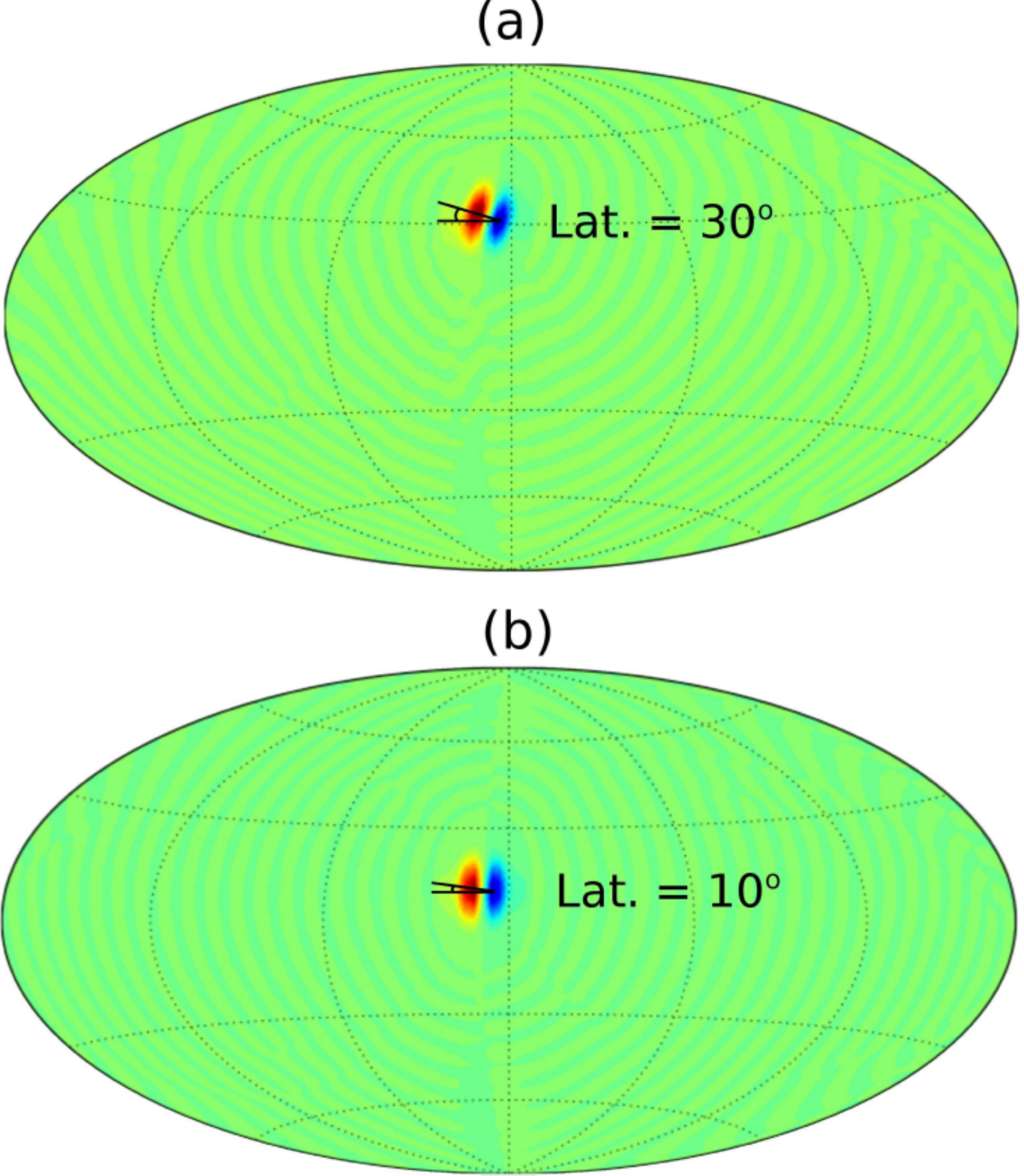}
\caption{Production of tilted bipolar regions at the surface: for higher latitude ($30^\circ$) the tilt angle is larger (subfigure a), whereas for lower latitude ($10^\circ$) it is smaller (subfigure b).}
\label{fig:tilt_vs_lat}
\end{figure}

The emergence of twisted magnetic flux ropes results in the production of tilted bipolar magnetic regions at the surface. The vortical velocity is tuned so that the tilt in BMRs is with respect to the east-west direction. Also, the tilt is clockwise in the northern hemisphere and anti-clockwise in the southern hemisphere where the tilt corresponds to values observed at the solar surface (between $4^\circ$ and $14^\circ$~\citep{Wang:SP1989}). The tilt in the bipolar regions is crucial for the Babcock-Leighton mechanism. We therefore tune the amplitudes $V^t_{\theta0}$ and $V^t_{\phi0}$ of the vortical velocity such that the produced BMRs obey Joy's law~\citep{Hale:APJ1919} for tilt angle and latitude. We do not introduce any fluctuations in the tilt angle of BMRs and then limit the variability of the magnetic solutions we obtain. In Fig.~\ref{fig:tilt_vs_lat}, we show one BMR emerging at two different latitudes where the BMR at a higher latitude ($30^\circ$) has a larger tilt angle as compared to the BMR at a lower latitude ($10^\circ$). In our model we allow $32$ BMRs to emerge every month. Observations suggest that on average around $100$ sunspots emerged every month during cycle $22$, around $80$ sunpots during cycle $23$, and around 60 sunspots during cycle $24$~\citep{sidc:be}. The total sunspots emerging during a sunspot cycle in our model is smaller than that observed in the Sun. However, the size of an individual spot in our case is large as compared to that of sunspots such that one individual spot in our model is equivalent to an observed sunspot group (containing 3-4 small-size spots). The selection of the sites of BMR emergence is random, as described earlier. The randomness in the selection of the sites of emerging BMRs is the only mechanism which can introduce variability in the dynamo solution. For example, if during half a magnetic cycle more BMRs emerge near the equator, then the leading spots of those BMRs would participate more in cross-equatorial cancellation, which would contribute more to the net poloidal field produced through the BL mechanism and hence the amplitude of that half cycle would be slightly higher. On the other hand, more BMRs emerging at higher latitudes would reduce the amplitude of a particular half cycle. However, if the number of emerging BMRs is large, then the BMRs at the solar surface would be evenly distributed (in the latitudinal region $[-35^\circ, +35^\circ]$), reducing the variability of the cycle amplitude.  

\section{Dynamo solution and the magnetic cycle}
\label{sec:dyn_sol_mag_cyc}  

In this section, we present the results of a typical dynamo simulation performed for the peak meridional flow $V_{\theta p} =12.5 \; \rm m \, \rm s^{-1}$ and convection zone diffusivity $\eta_c=7 \times 10^{10} \, \rm cm^2 \, \rm s^{-1}$. The initial condition and the other velocity and diffusivity parameters are the ones described in Sect.~\ref{sec:fl_tr_mdel}.  
       
\begin{figure}[htbp]
\centering
\includegraphics[scale=0.37]{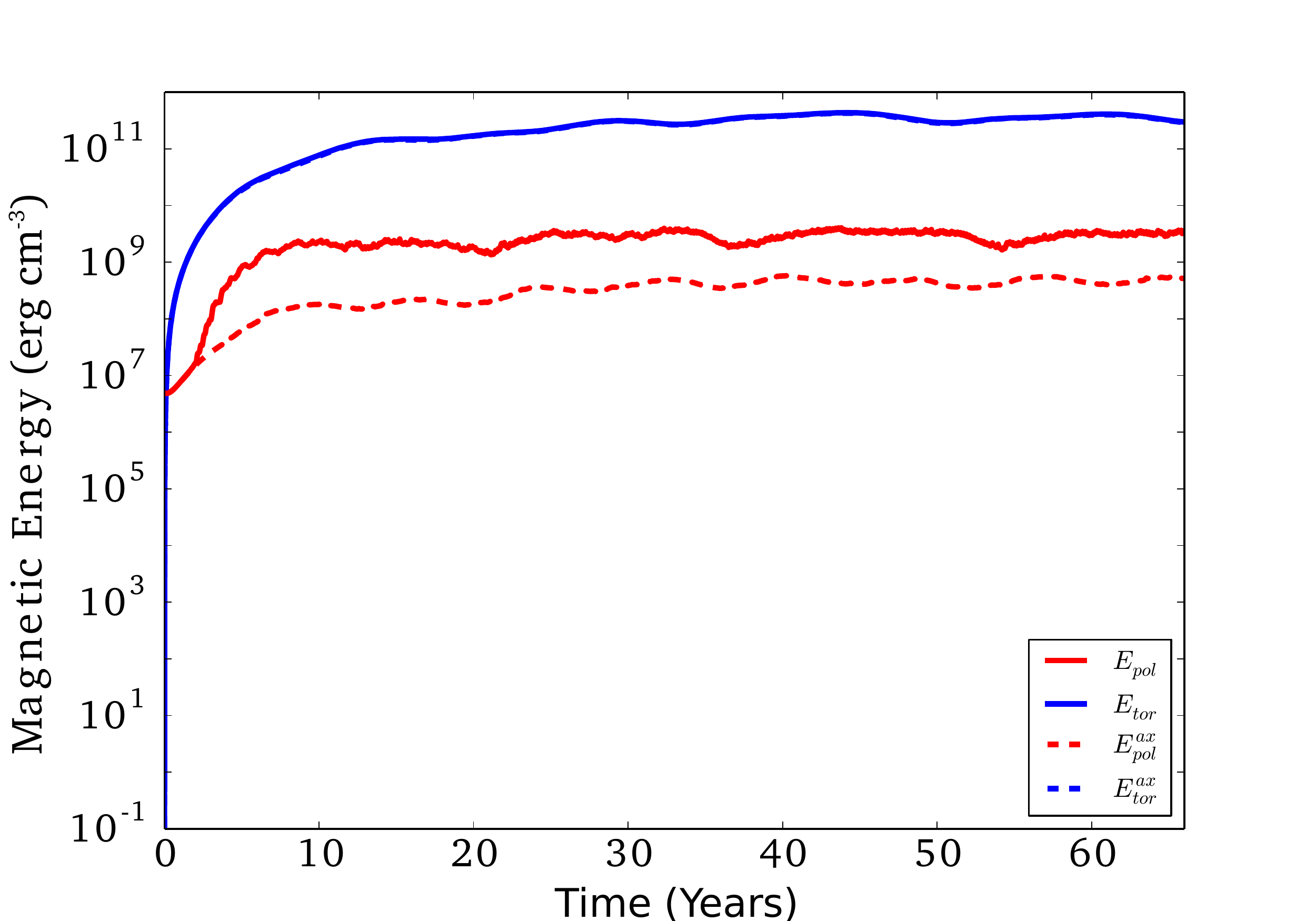}
\caption{The time-evolution of poloidal and toroidal components of the magnetic energy ($E_{pol}$ and $E_{tor}$, respectively) showing a self-sustained saturated dynamo in the system. Here $E_{pol}^{ax}$ and $E_{tor}^{ax}$ represent the axisymmetric parts of poloidal and toroidal energies, respectively. The poloidal energy is mainly dominated by its nonaxisymmetric part.}
\label{fig:mag_en_mc_12}
\end{figure}

In Fig.~\ref{fig:mag_en_mc_12}, we illustrate the time-evolution of the mean toroidal and the mean poloidal magnetic energies, which shows an initial exponential growth of the magnetic energy followed by the saturation of energy. We note that the poloidal energy is dominated by its nonaxisymmetric part, which is mainly due to the emergence of tilted BMRs that are nonaxisymmetric in nature. The magnetic energy attains saturation due to the lower and upper cutoffs on the buoyantly emerging toroidal field. These cutoffs limit the magnetic flux reaching the surface and act as the quenching effect in this kinematic dynamo. 

\begin{figure*}[htbp]
\centering
\includegraphics[scale=0.45]{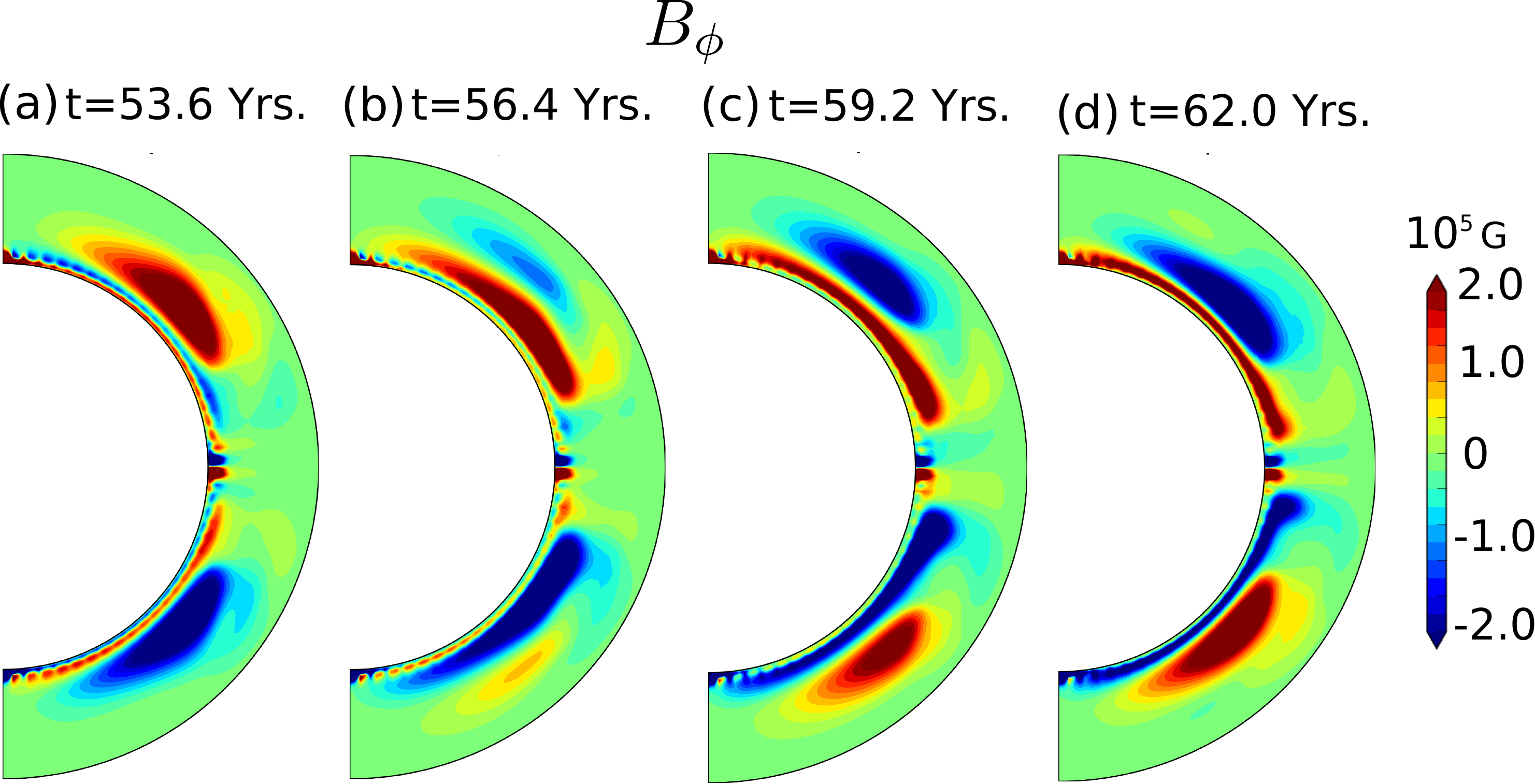}
\caption{Snapshots of the longitudinally averaged toroidal magnetic field ($B_\phi$) at different stages during half a magnetic cycle. Subfigures show the polarity reversals of the field in the two hemispheres.}
\label{fig:bp_avg}
\end{figure*}

\begin{figure*}[htbp]
\centering
\includegraphics[scale=0.45]{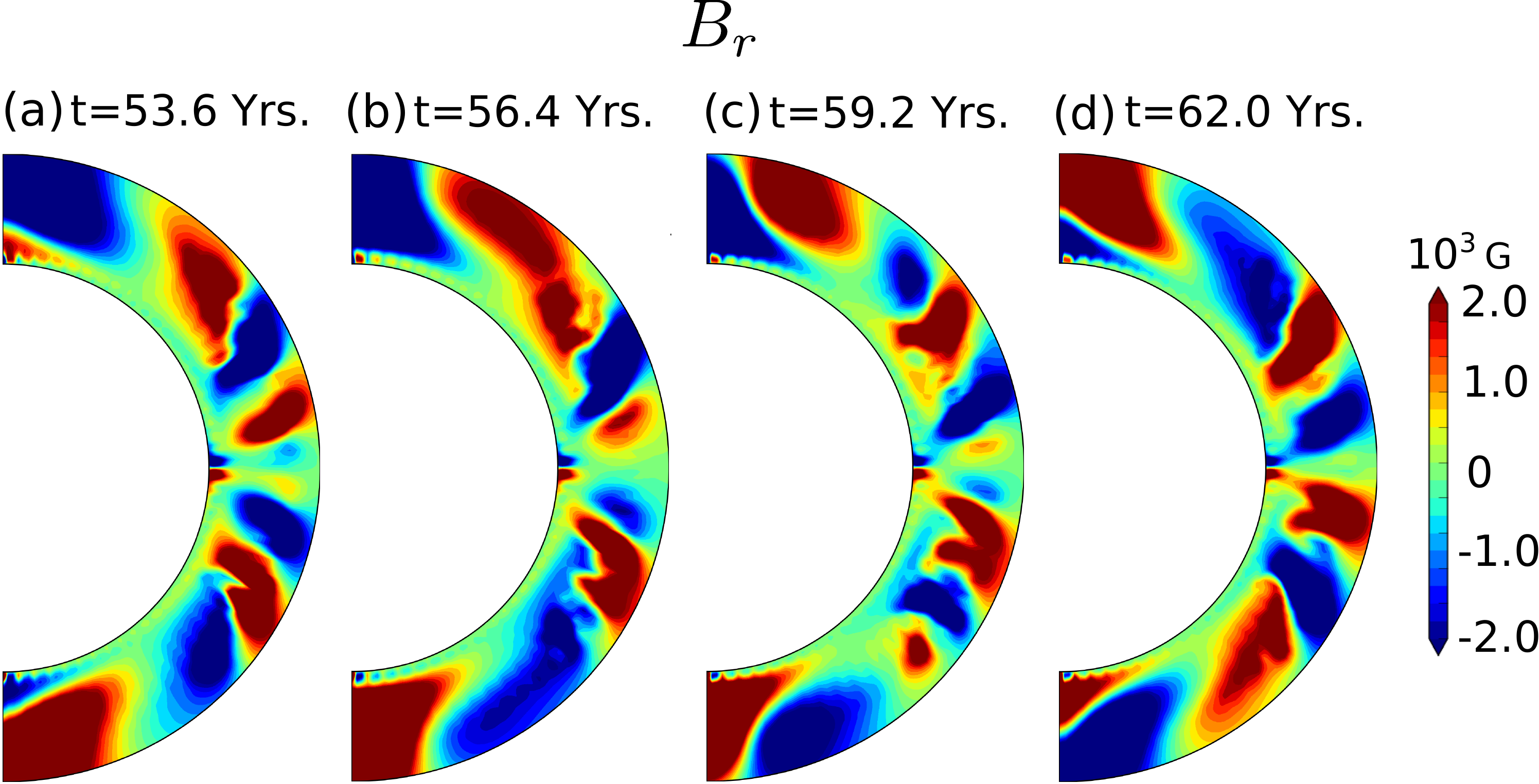}
\caption{Snapshots of the longitudinally averaged radial magnetic field ($B_r$) at different stages during half a magnetic cycle. Subfigures show the polarity reversals of the polar field.}
\label{fig:br_avg}
\end{figure*}

In Fig.~\ref{fig:bp_avg}, we show the snapshots of longitudinally averaged toroidal field during half a magnetic cycle. At the beginning of the magnetic cycle, the northern hemisphere convection zone is dominated by a strong toroidal field with a positive polarity [see Fig.~\ref{fig:bp_avg}(a)]. Afterwards, the toroidal field diffuses and gets advected by the meridional flow towards the equator [see Fig.~\ref{fig:bp_avg}(b)], which then becomes buoyantly transported to the surface to produce BMRs. Later, we observe the production of a toroidal field (with negative polarity) in the subsequent cycle [see Fig.~\ref{fig:bp_avg}(c)], which is again subject to advection by the meridional flow towards the equator [see Fig.~\ref{fig:bp_avg}(d)]. Figure~\ref{fig:br_avg} illustrates the snapshots of the longitudinally averaged radial component of the (poloidal) magnetic field at the same instances as the toroidal field of Fig.~\ref{fig:bp_avg}. At the beginning of the magnetic cycle, the dominant positive toroidal field in the convection zone (see northern hemisphere of Fig.~\ref{fig:bp_avg}a) produces BMRs with positive trailing spots. Hence in the northern hemisphere a net positive poloidal flux is produced and advected towards the poles (see Fig.~\ref{fig:br_avg}(a,b)), to end up with the reversal of the polar field of the previous cycle (see Fig.~\ref{fig:br_avg}(d)). Later, the dominant toroidal field is of negative polarity (see northern hemisphere of Fig.~\ref{fig:bp_avg}(d)), which produces BMRs with negative trailing spots, which then results in a net negative poloidal flux being produced and transported towards the pole to initiate the next polar field reversal (see Fig.~\ref{fig:br_avg}(d)). We note that the small-scale numerical artefacts appearing in Figs. ~\ref{fig:bp_avg} and ~\ref{fig:br_avg} at the very
bottom of the simulation domain do not affect the overall dynamics, which is initiated and mostly limited to the convection zone. Expensive numerical computation time limits our ability to explore much-higher-resolution simulations in this exhaustive parameter space study.

\begin{figure}[htbp]
\centering
\includegraphics[scale=0.45]{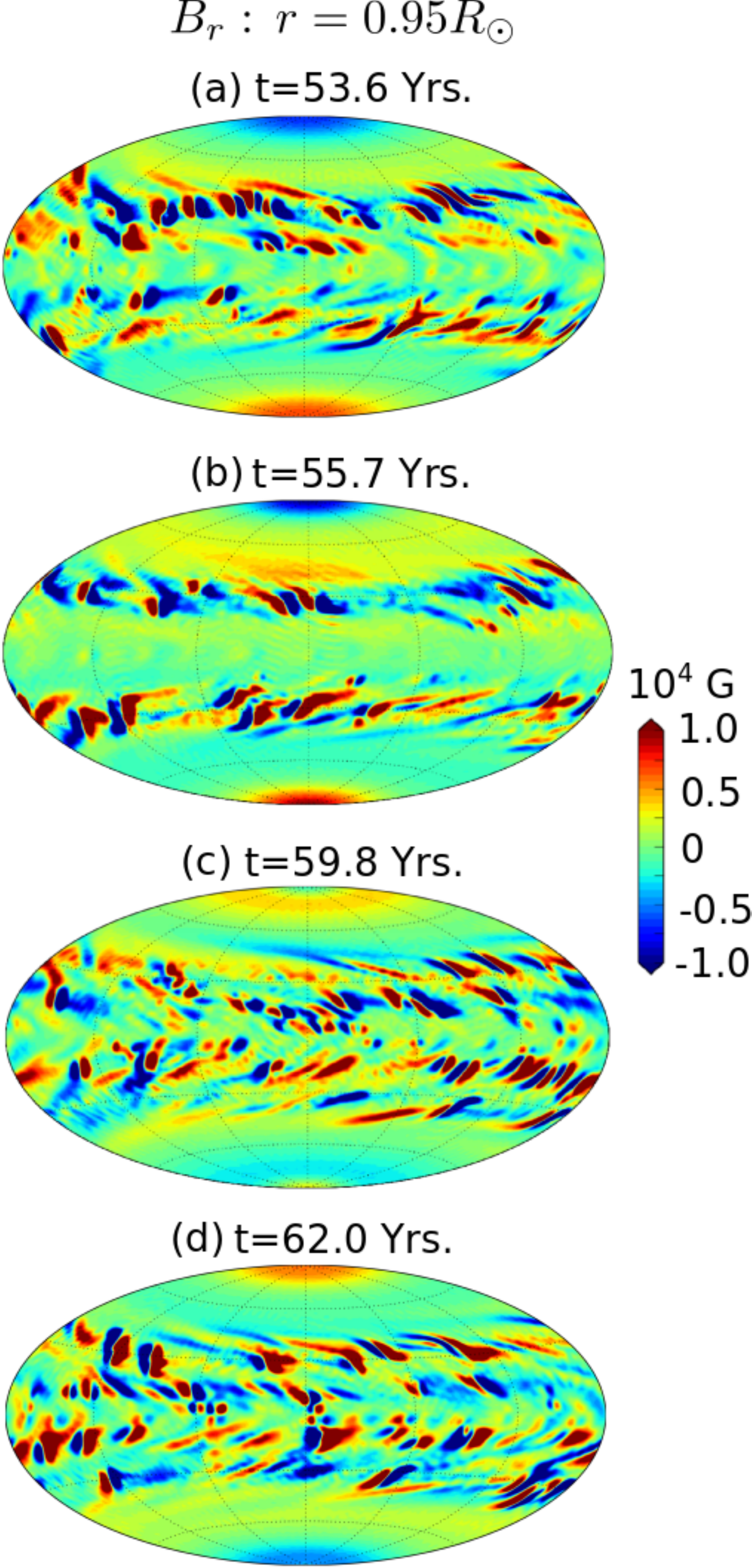}
\caption{Snapshots of the radial magnetic field ($B_r$) at the surface showing tilted BMRs and large-scale polar magnetic field (a) before reversals, (b) during sunspot minimum, (c) during sunspot maximum and field reversals, and (d) after the polarity reversals through the BL mechanism.}
\label{fig:bmr_rev}
\end{figure}

In Fig.~\ref{fig:bmr_rev}(a), the surface magnetic field ($B_r$) shows a large number of bipolar spots and the presence of a large-scale polar cap. In the northern hemisphere, the BMRs with positive trailing polarity seen at mid-latitudes are produced by the positive toroidal field in the convection zone (see Fig.~\ref{fig:bp_avg}a). If we take the longitudinal average of the surface magnetic field, then due to the tilt of BMRs, we obtain a nonzero magnetic flux of the polarity of the trailing spots, which is clearly visible at mid and high latitudes in both hemispheres. In Fig.~\ref{fig:bmr_rev}(b) we see a relatively small number of BMRs at the surface close to the equator, which represents here a sunspot minimum. We note however that in this case we do not obtain a proper sunspot minimum with only a few BMRs at the surface. Indeed in our model, we observe overlaps between two consecutive cycles which results in a situation where BMRs of the previous cycle linger at the surface along with the BMRs of the next cycle. This is one of the shortcomings of our dynamo model and future work will be devoted to finding the appropriate parameters enabling  this cycle overlap to be reduced. At a later stage, we observe a large number of tilted BMRs emerging at the surface (see Fig.~\ref{fig:bmr_rev}c), which produce a significant amount of net magnetic flux of polarity opposite to that of the polar field. The net magnetic flux then gets advected, due to surface flows, towards the poles to reverse the polar field, which is precisely the Babcock-Leighton mechanism. The polar field reversal happens near the sunspot maximum, consistent with the solar-cycle observations. Later, the newly generated poloidal field produces a toroidal field of polarity opposite to the previous one (see Fig.~\ref{fig:bp_avg}d), which then produces BMRs of opposite polarity (negative trailing spots  in the northern hemisphere), as shown in Fig.~\ref{fig:bmr_rev}(d) at mid-latitudes. The polarities of produced bipolar spots change along with the polarity of the toroidal field in the convection zone. 

One of the shortcomings of our dynamo model is that the strength of an individual BMR and hence the polar field strength are quite high ($10^3$ gauss) as compared to those observed in the Sun (polar field strength $\approx 2$ gauss~\citep{Jaramillo:APJ2012}). We note that in our model, the toroidal fields of magnitude ($4 \times 10^4, 1.4 \times 10^5$) gauss are subject to magnetic buoyancy in the convection zone and emerge at the surface to produce BMRs. These emerging toroidal fluxes, after diffusion in the convection zone, produce BMRs with a  field strength of the order of a kilogauss. If we increase the lower cutoff or decrease the upper cutoff on buoyantly emerging $B_{\phi}$ significantly, then it results in the continuous decay of the magnetic field, due to a high surface magnetic diffusivity, killing the dynamo in the system. Therefore, we keep the aforementioned cutoff on the emerging toroidal field in order to obtain a self-sustained dynamo. \cite{Miesch:ASR2016} have also reported strong magnetic fields in their dynamo model. In our model, the magnetic flux of an individual spot is of the order of $10^{22}$ Mx, which is quite high compared to that of an individual sunspot ($10^{20} - 5 \times 10^{21}$ Mx~\citep{Cheung:APJ2010}). A possible explanation for such high polar fields in kinematic dynamo simulations and their consistency with observations of unipolar kilogauss flux tubes in the polar regions is alluded to in \cite{Nandyetal2011}.

\begin{figure}[htbp]
\centering
\includegraphics[scale=0.4]{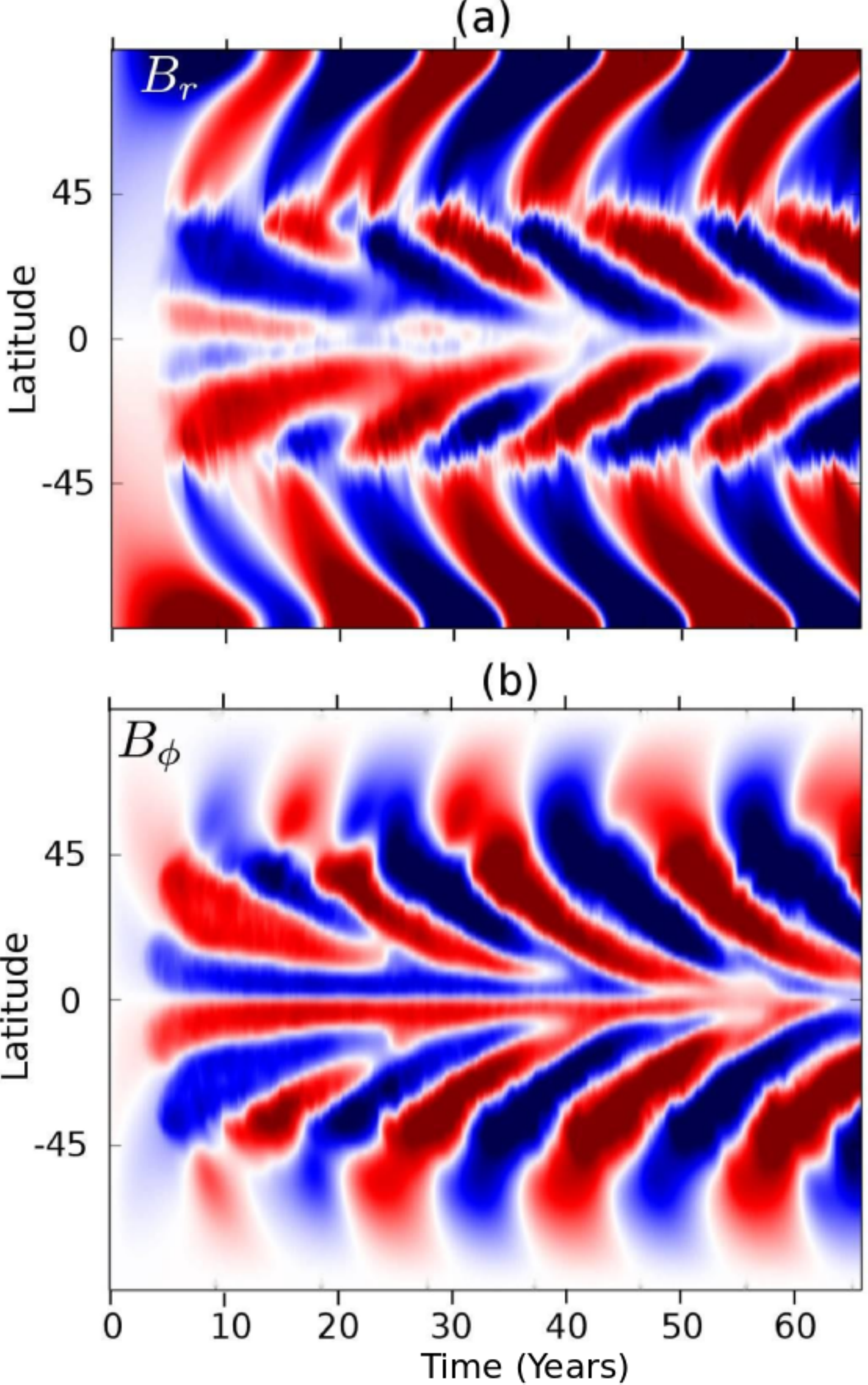}
\caption{Maps of the mean radial magnetic field ($B_r$) at the surface and the mean toroidal field ($B_{\phi}$) at the base of the convection zone showing cyclic field reversals. The toroidal field gets advected towards the equator with time (b) due to the meridional circulation in the convection zone.}
\label{fig:butterfly}
\end{figure}

\begin{figure}[htbp]
\centering
\includegraphics[scale=0.085]{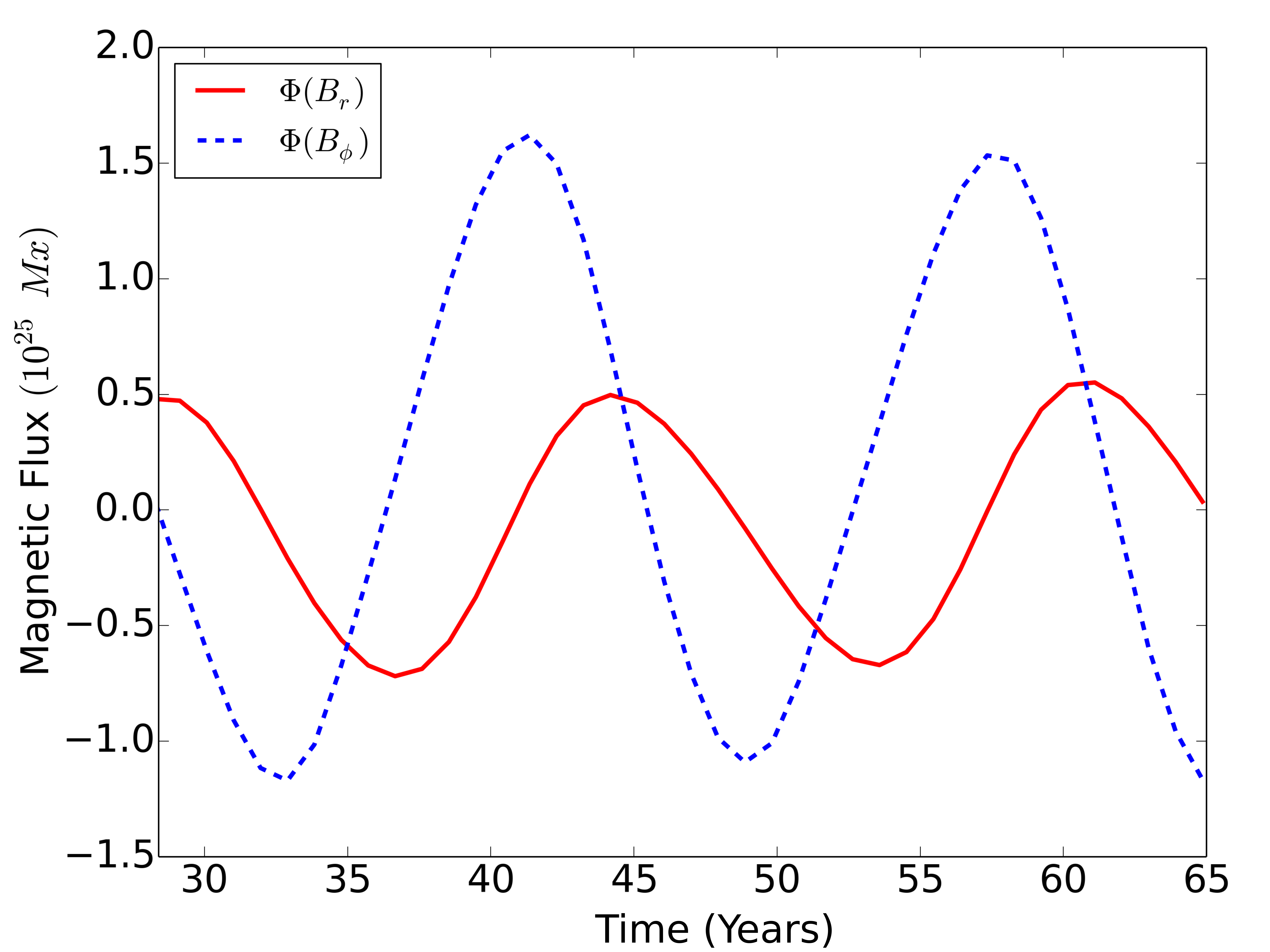}
\caption{The time-evolution of the poloidal magnetic flux [$\Phi(B_r)$] and the toroidal magnetic flux [$\Phi(B_\phi)$] showing periodic field reversals. There is a $90^\circ$ phase difference between the toroidal and poloidal fluxes.}
\label{fig:flux_Br_Bphi_t}
\end{figure}

We plot maps of the mean radial magnetic field at the surface (Fig.~\ref{fig:butterfly}a) and the mean toroidal field at the base of the convection zone (Fig. ~\ref{fig:butterfly}b), which shows regular polarity reversals of the polar field as well as the toroidal field. The polar field reversals happen near the peak of the toroidal field in the convection zone that corresponds to maximum of BMRs at the surface. The average time for the polar field reversals is $8.5$ years. As we see in the following section, the duration of half a magnetic cycle is highly sensitive to the meridional flow amplitude, we then do get an $11$-year cycle (as observed in the Sun) when we choose the peak meridional flow speed to be $7.5 \, \rm m \, \rm s^{-1}$. We estimate the amplitude of the magnetic cycle by computing the poloidal magnetic flux $\Phi(B_r)$ of the polar cap (at $r = 0.95 \, R_\odot$), and the toroidal magnetic flux $\Phi(B_\phi)$ near the tachocline; which are defined as
\begin{eqnarray}
\Phi(B_r) &=& \int_{\theta = 0^\circ}^{20^\circ} \int_{\phi = 0}^{2 \pi} B_r \, r^2 \, \sin\theta \, d \theta \, d \phi,  \\
\Phi(B_\phi) &=& \int_{\theta = 40^\circ}^{90^\circ} \int_{r = 0.7 \, R_\odot}^{0.725 \, R_\odot} B_\phi \, r \, d \theta \, dr,
\end{eqnarray}
where $\theta$ represents the colatitudes. In Fig.~\ref{fig:flux_Br_Bphi_t}, we illustrate the time-evolution of the amplitude of the magnetic cycle. The toroidal flux peaks when the poloidal flux is minimum, and vice versa. Hence the toroidal and poloidal fluxes are in anti-phase, which is consistent with the solar observations~\citep{Hathaway:LRSP2010} where the sunspot number-strength is a manifestation of the strength of the toroidal flux in the convection zone. In our simulation, we observe a small degree of variability in the magnetic cycle, that is, the cycle amplitude varies for different cycles. In our model, BMRs emerge at random longitudes and latitudes, which may contribute different amounts of magnetic flux to the net surface magnetic flux generated through the BL process, and hence the amplitude of the magnetic cycle may vary accordingly. However, the level of variability will be small compared to simulations where the tilt angle is also allowed to fluctuate around a mean value. Further studies with this model would include tilt angle fluctuations and anti-Hale regions which are shown to have significant impact on cycle amplitudes in 2D surface flux-transport models~\citep{Nagy:SP2017}.

\section{Parameter-space study}
\label{sec:para_study}

In the previous section, we discussed the results of a typical (i.e., our standard) dynamo simulation which was obtained by fixing various parameters such as the peak meridional flow speed, the convection zone diffusivity, and the frequency of BMR emergence. In this section, we vary these parameters and study their effects on the magnetic cycle to explore solar cycle dynamics under the dominance of diverse flux-transport regimes.  

\subsection{Effect of meridional circulation and convection zone diffusivity on cycle duration}

We examine the effect of meridional flow on the magnetic cycle by changing the peak value of the meridional flow speed in the convection zone, keeping the flow profiles the same as shown in Fig.~\ref{fig:diff_mer_fl_num}. For this purpose, we chose the peak meridional flow speed to be $V_{\theta p} = 7.5, 10, 12.5, 15, 20 \, \rm m \rm s^{-1}$. 

\begin{figure}[htbp]
\centering
\includegraphics[scale=0.3]{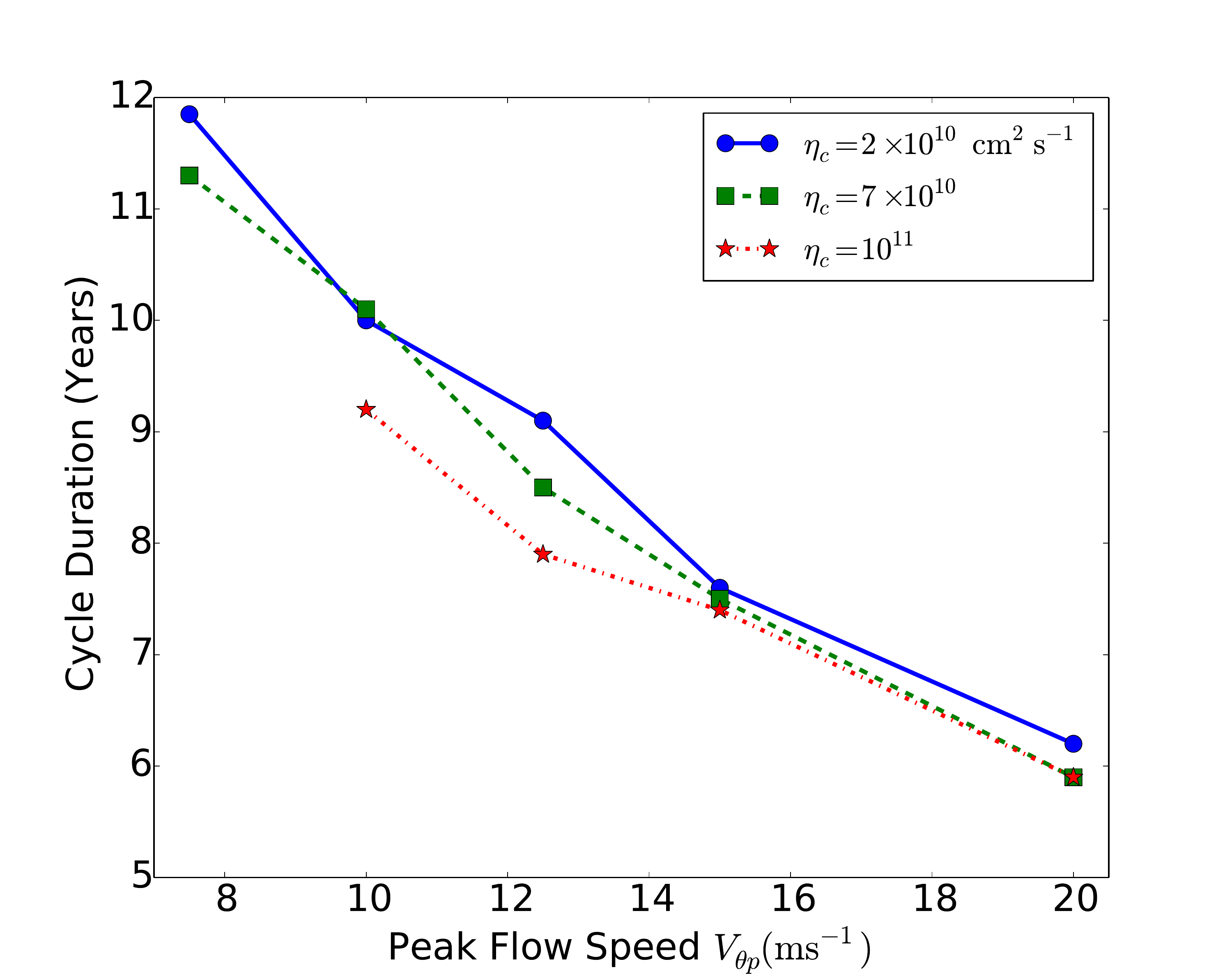}
\caption{The duration of half a magnetic cycle ($T_{1/2}$) with peak meridional flow speed ($V_{\theta p}$) at different values of the convection zone diffusivity ($\eta_c$). Length of half a magnetic cycle decreases with increasing $V_{\theta p}$ for all $\eta_c$.}
\label{fig:t_half_mc}
\end{figure}

For a fixed convection zone diffusivity, we observe that the cycle duration is highly sensitive to the peak meridional flow speed. The duration of half a magnetic cycle, that is, the sunspot cycle ($T_{1/2}$) decreases with the peak meridional flow speed following the relationship: $T_{1/2} \propto V_{\theta p}^{-0.67}$ (here $T_{1/2}$ is in years and $V_{\theta p}$ is in meters per second) for the convection zone diffusivity $\eta_c = 2 \times 10^{10} \, \rm cm^2 \, \rm s^{-1}$, and $T_{1/2} \propto V_{\theta p}^{-0.62}$ for $\eta_c = 10^{11} \, \rm cm^2 \, \rm s^{-1}$. This suggests that for a large $\eta_c$ the cycle duration depends less on the meridional flow, because in this case the diffusion plays an important role in the transport of magnetic flux. In Fig.~\ref{fig:t_half_mc}, we plot the length of half a magnetic cycle (the time duration between two consecutive polarity reversals) as a function of $V_{\theta p}$ at different values of $\eta_c$. At a fixed $V_{\theta p}$, $T_{1/2}$ slightly decreases as we increase $\eta_c$. The decrease is not very significant for large $V_{\theta p}$, but it is higher for the smaller values of $V_{\theta p}$. This is explained by the fact that for large $V_{\theta p}$ the flux-transport in the convection zone is mostly advection driven and cycle duration is mainly governed by the meridional circulation. On the other hand, for small values of $V_{\theta p}$, the diffusion also contributes to flux-transport in the convection zone, which in turn affects the cycle duration in a significant way.    

\subsection{Effect of meridional circulation and convection zone diffusivity on cycle amplitude}

\begin{figure}[htbp]
\centering
\includegraphics[scale=0.3]{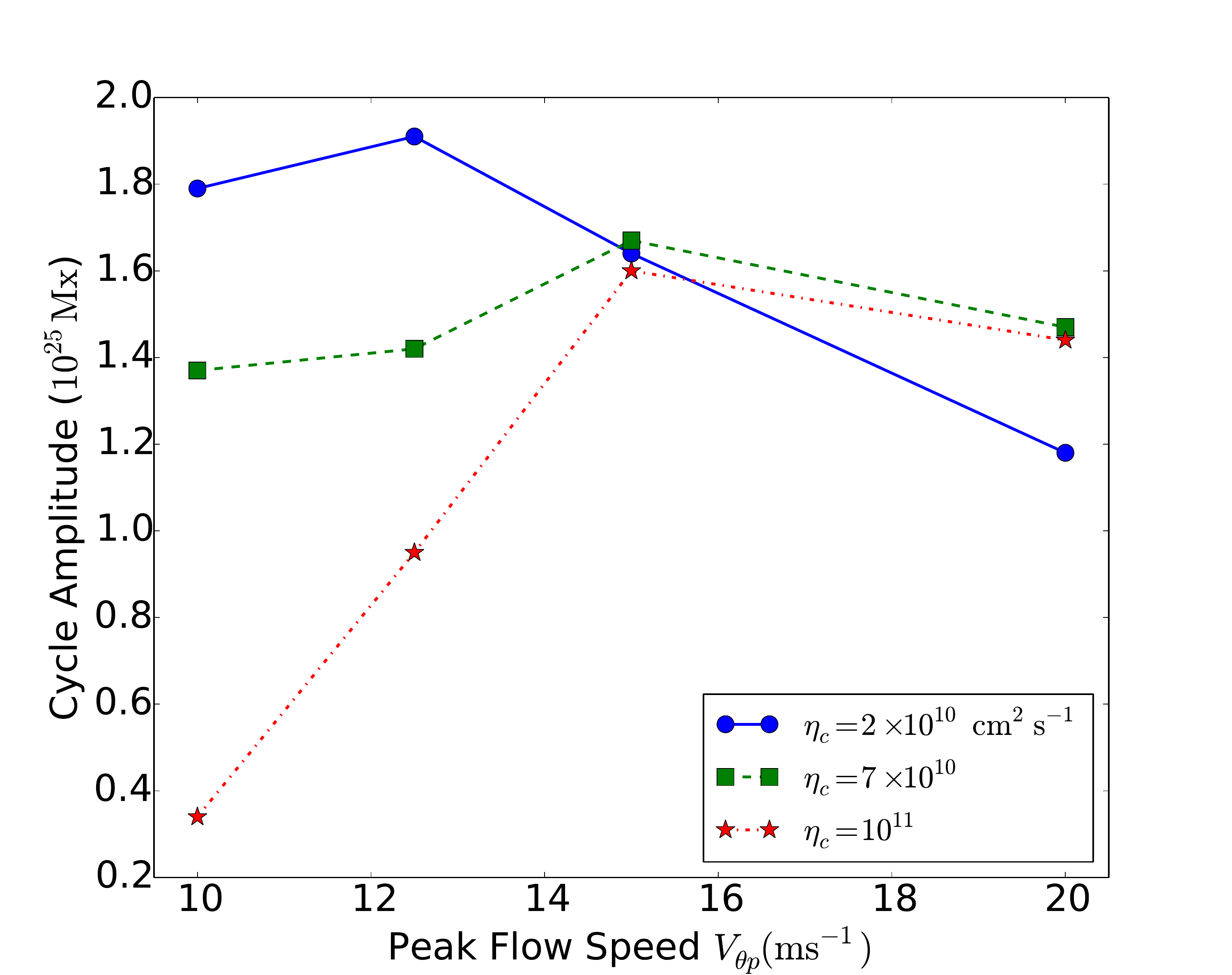}
\caption{Dependence of the cycle amplitude on the peak meridional flow speed ($V_{\theta p}$) at different values of the convection zone diffusivity ($\eta_c$). The cycle amplitude first increases and then decreases with increasing $V_{\theta p}$.}
\label{fig:fl_tor_mc}
\end{figure}

\begin{figure}[htbp]
\centering
\includegraphics[scale=0.3]{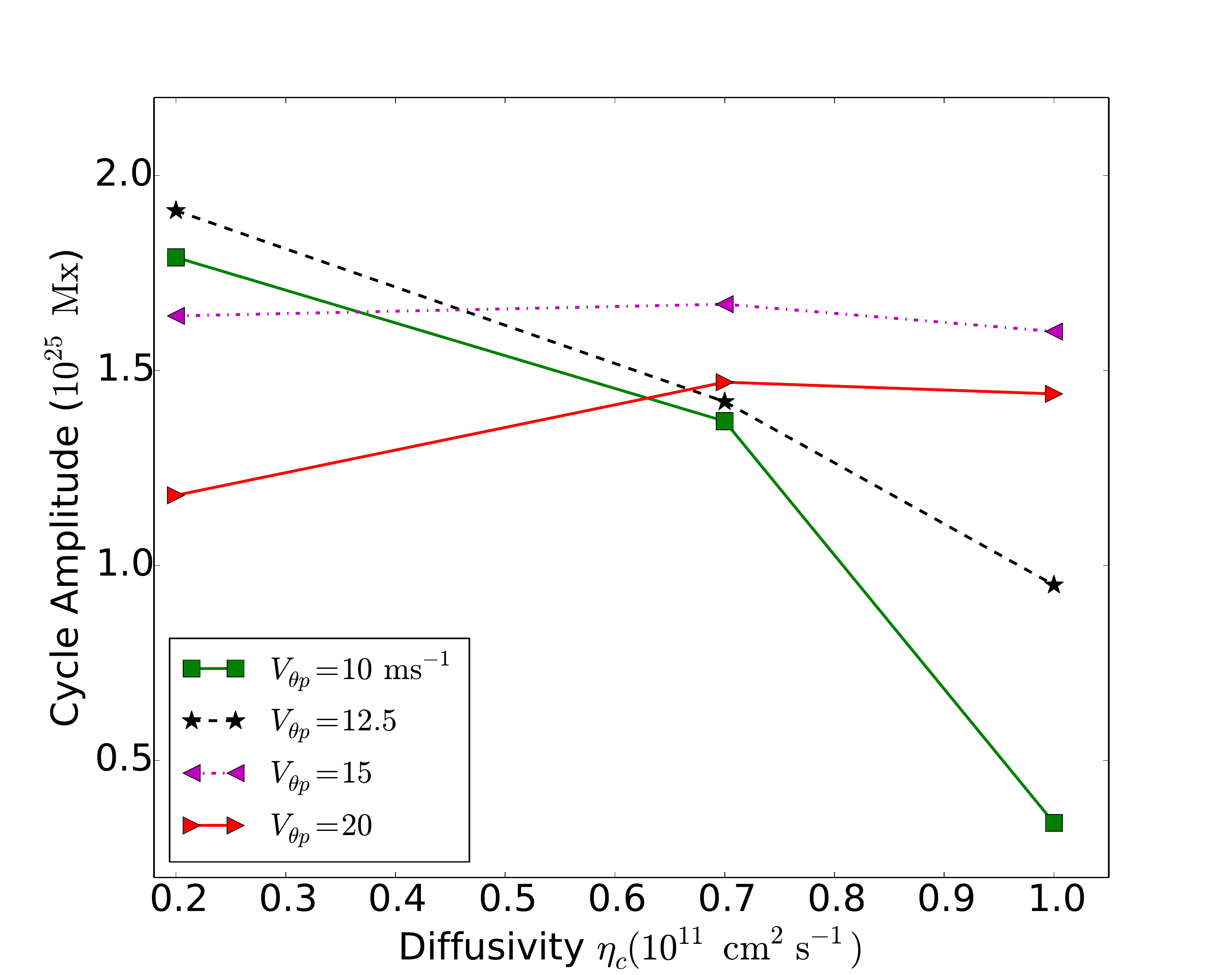}
\caption{Dependence of the cycle amplitude on the convection zone diffusivity ($\eta_c$) at different peak meridional flow speeds ($V_{\theta p}$). For high peak meridional flow speed ($V_{\theta p} =15, 20 \; \rm m \, \rm s^{-1}$ ), the cycle amplitude first increases and then decreases with increasing $\eta_c$, whereas for low peak meridional flow speed ($V_{\theta p} =10, 12.5 \; \rm m \, \rm s^{-1}$), the cycle amplitude continuously decreases with increasing $\eta_c$.}
\label{fig:fl_tor_diff}
\end{figure}

To understand the impact of meridional circulation and convection zone diffusivity on the cycle amplitude, we compute the toroidal magnetic flux $\Phi(B_\phi)$ (discussed in Sect.~\ref{sec:dyn_sol_mag_cyc}). In Fig.~\ref{fig:fl_tor_mc}, we illustrate the cycle amplitude as a function of peak meridional flow speed at different values of $\eta_c$. The toroidal flux plotted in the figure represents the averaged peak values for several magnetic cycles in the steady state of the dynamo run. We observe that for a fixed $\eta_c$ the cycle amplitude first increases and then decreases with $V_{\theta p}$. In Fig.~\ref{fig:fl_tor_diff}, we plot the cycle amplitude as a function of $\eta_c$ at different values of $V_{\theta p}$. For low meridional flows ($V_{\theta p} =10, 12.5 \; \rm m\,\rm s^{-1}$), the cycle amplitude decreases with increasing $\eta_c$. On the other hand, for high peak meridional flows ($V_{\theta p} =15, 20 \; \rm m\,\rm s^{-1}$), the cycle amplitude first increases and then decreases with increasing $\eta_c$. 

To explain the aforementioned trends we follow \cite{Yeates:APJ2008} and define a low diffusivity and high meridional circulation speed regime as the advection-dominated regime, and a high diffusivity and low meridional circulation speed regime as the diffusion-dominated regime. In the advection-dominated regime (low $\eta_c$ and high $V_{\theta p}$), we observe a lower cycle amplitude as the circulation speed is increased (see Fig.~\ref{fig:fl_tor_mc}), because a high circulation speed allows less time for toroidal field to be amplified near the tachocline. In this regime, for a fixed meridional speed, the cycle amplitude increases with increasing $\eta_c$ (see Fig.~\ref{fig:fl_tor_diff}). This kind of trend is observed because of a significant direct diffusive flux-transport of the poloidal field across the convection zone, as suggested by \cite{Yeates:APJ2008}. In the diffusion-dominated regime (high $\eta_c$ and low $V_{\theta p}$), the cycle amplitude increases with the circulation speed and the increase is more significant for high $\eta_c$ (see Fig.~\ref{fig:fl_tor_mc}). In this regime when the circulation speed is increased, the time available for diffusive decay of the poloidal field being transported in the convection zone is less, which then allows the production of higher toroidal field generated through shearing of stronger poloidal field.

\subsection{Evolution of the magnetic field in the advection- and diffusion-dominated regimes}

\begin{figure}[htbp]
\centering
\includegraphics[scale=0.35]{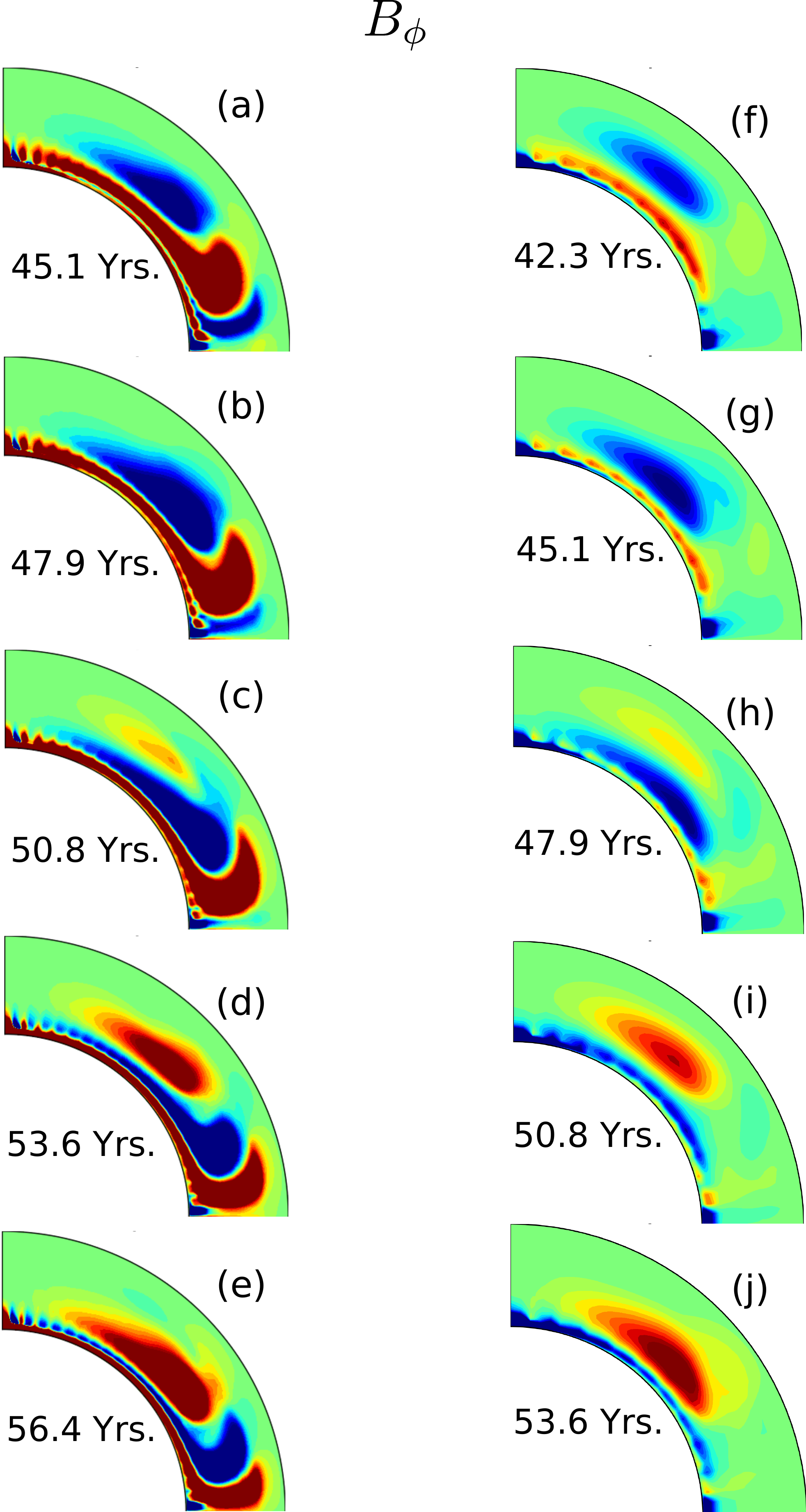}
\caption{Snapshots of the longitudinally averaged toroidal magnetic field ($B_\phi$) at different stages during half a magnetic cycle, with peak meridional circulation $V_{\theta p} =12.5 \, \rm m  \, \rm s^{-1}$. The left column represents the evolution of the toroidal field for the advection-dominated convection zone ($\eta_c=2 \times 10^{10} \, \rm cm^2 \, \rm s^{-1}$) where the field is significantly advected in the convection zone. The right column corresponds to the diffusion-dominated convection zone ($\eta_c=10^{11} \, \rm cm^2 \, \rm s^{-1}$) where the diffusion of the field is significant. In the advection-dominated case, the toroidal fields of previous few cycles are also present near the tachocline.}
\label{fig:tor_field_comp}
\end{figure}

Here we study the evolution of toroidal and poloidal magnetic fields in the advection- and diffusion-dominated regimes. In order to distinguish between advection- and diffusion-dominated regimes, we compute the ratio ($\rm Re_M$) of the diffusion timescale ($\tau_{Diff} = L_c^2/\eta_c$, where $L_c = 0.3 \, R_\odot$ is the radial distance across the convection zone) and the advection time-scale ($\tau_{Adv} = \frac{\pi R_{\odot}}{2 V_{\theta p}}$, the time taken by the meridional flow to advect a fluid particle from the equator to the pole at the surface), which indicates that above (below) $\rm Re_M \approx 75$ we have an advection-dominated (diffusion-dominated) regime. In Fig.~\ref{fig:tor_field_comp}, we illustrate the time evolution of the toroidal field during half a magnetic cycle for $V_{\theta p} =12.5 \, \rm m \, \rm s^{-1}, \eta_c=2 \times 10^{10} \, \rm cm^2\, \rm s^{-1}$ ($\rm Re_M = 246$, advection-dominated convection zone: Fig.~\ref{fig:tor_field_comp}a-e) and for $V_{\theta p} =12.5 \, \rm m \, \rm s^{-1}, \eta_c=10^{11} \, \rm cm^2 \, \rm s^{-1}$ ($\rm Re_M = 49$, diffusion-dominated convection zone: Fig.~\ref{fig:tor_field_comp}f-j). In an advection-dominated regime, the toroidal field generated in the convection zone gets advected, by the meridional circulation, towards the equatorial region. On the other hand, in a diffusion-dominated regime, the generated toroidal field gets diffused quickly before being advected by the meridional flow. In an advection-dominated regime, when the toroidal field with positive polarity gets amplified in the convection zone (cycle $n$), we observe the residual toroidal field of cycle $n-1$ (negative polarity) near the tachocline, and below that a thin layer of a positive toroidal field of cycle $n-2$ (see Fig.~\ref{fig:tor_field_comp}c,d). This suggests that the toroidal fields of cycles $n$, $n-1$, and $n-2$ participate in the production of bipolar regions of cycle $n$. Therefore the advection-dominated case produces magnetic cycles with memories of the previous two cycles. On the other hand, in the diffusion-dominated regime, the toroidal field quickly diffuses, leaving a weak residual toroidal field of only cycle $n-1$ (see Fig.~\ref{fig:tor_field_comp}i), which suggests that the magnetic cycle retains memory of only the previous cycle. This is in agreement with previous studies such as \cite{Yeates:APJ2008}.

\begin{figure}[htbp]
\centering
\includegraphics[scale=0.35]{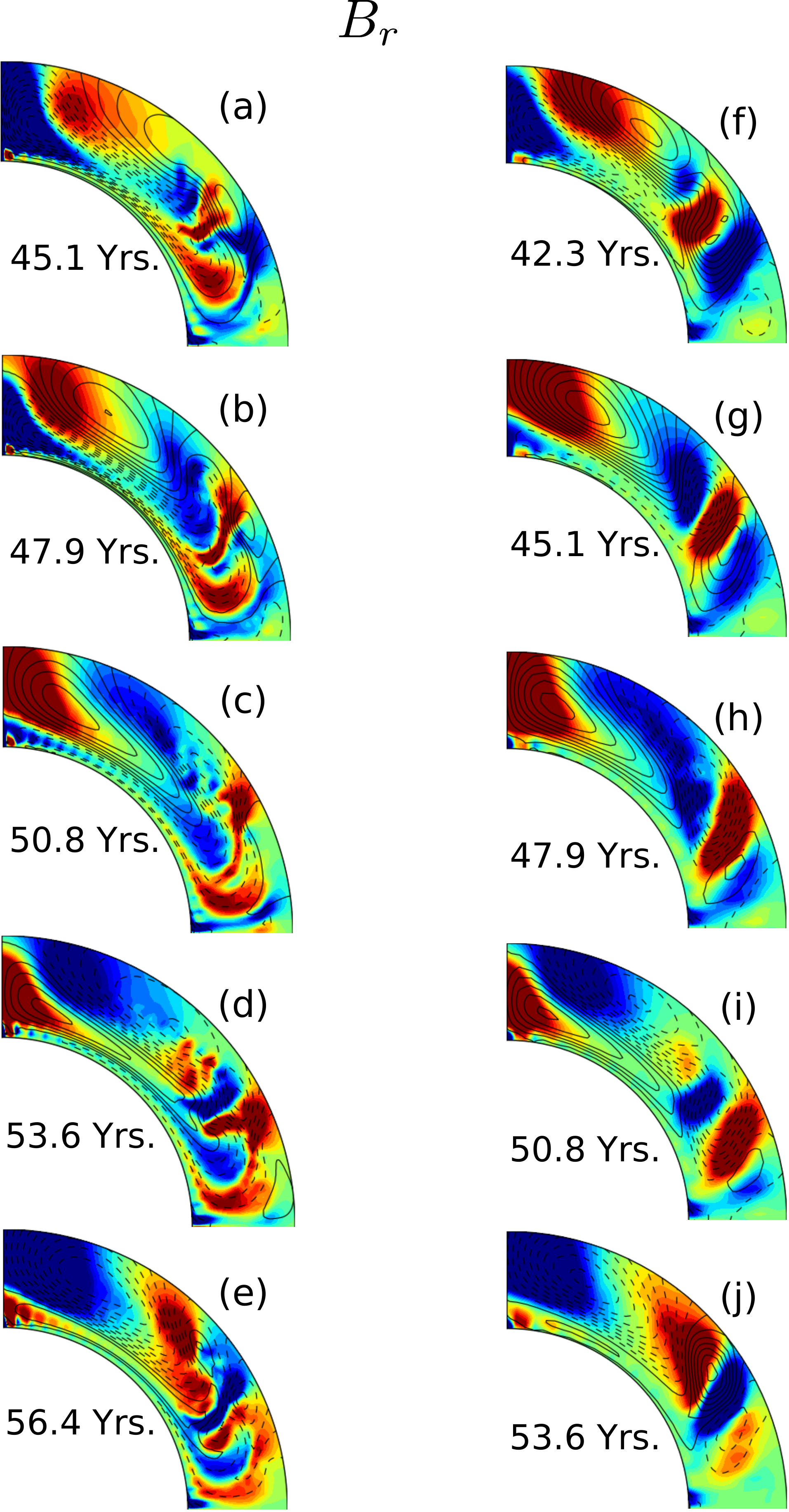}
\caption{Snapshots of the longitudinally averaged radial magnetic field ($B_r$) at different stages during half a magnetic cycle, with peak meridional circulation $V_{\theta p} =12.5 \, \rm m \, \rm s^{-1}$. The left column illustrates the evolution of the poloidal field for the advection-dominated convection zone ($\eta_c=2 \times 10^{10} \, \rm cm^2 \, \rm s^{-1}$), whereas the right column is for the diffusion-dominated convection zone ($\eta_c=10^{11} \, \rm cm^2 \, \rm s^{-1}$). The solid lines represent clockwise, while the dashed lines anti-clockwise poloidal fields, respectively.}
\label{fig:pol_field_comp}
\end{figure}

Figure~\ref{fig:pol_field_comp} illustrates the evolution of the poloidal field during half a magnetic cycle for the advection-dominated (Fig.~\ref{fig:pol_field_comp}a-e) and diffusion-dominated cases (Fig.~\ref{fig:pol_field_comp}f-j). At an instance when a clockwise poloidal field of cycle $n$ dominates, we observe layers of the anti-clockwise field of cycle $n-1$ near the tachocline, and below that a very thin layer of clockwise field of cycle $n-2$ (see Fig.~\ref{fig:pol_field_comp}b). This shows that the toroidal field of cycle $n+1$ gets produced by the poloidal fields of the previous few cycles. For the diffusion-dominated system, however, we do not observe many different layers of the poloidal field. The time-evolution of poloidal field is in agreement with the understanding that for an advection-dominated convection zone the memories of the previous few cycles propagate to the subsequent cycle. The propagation of memories of earlier cycles plays an important role in determining the amplitude of the future sunspot cycle and is therefore important for solar-cycle predictions. 

\subsection{Effect of number of emerging BMRs}

\begin{figure}[htbp]
\centering
\includegraphics[scale=0.8]{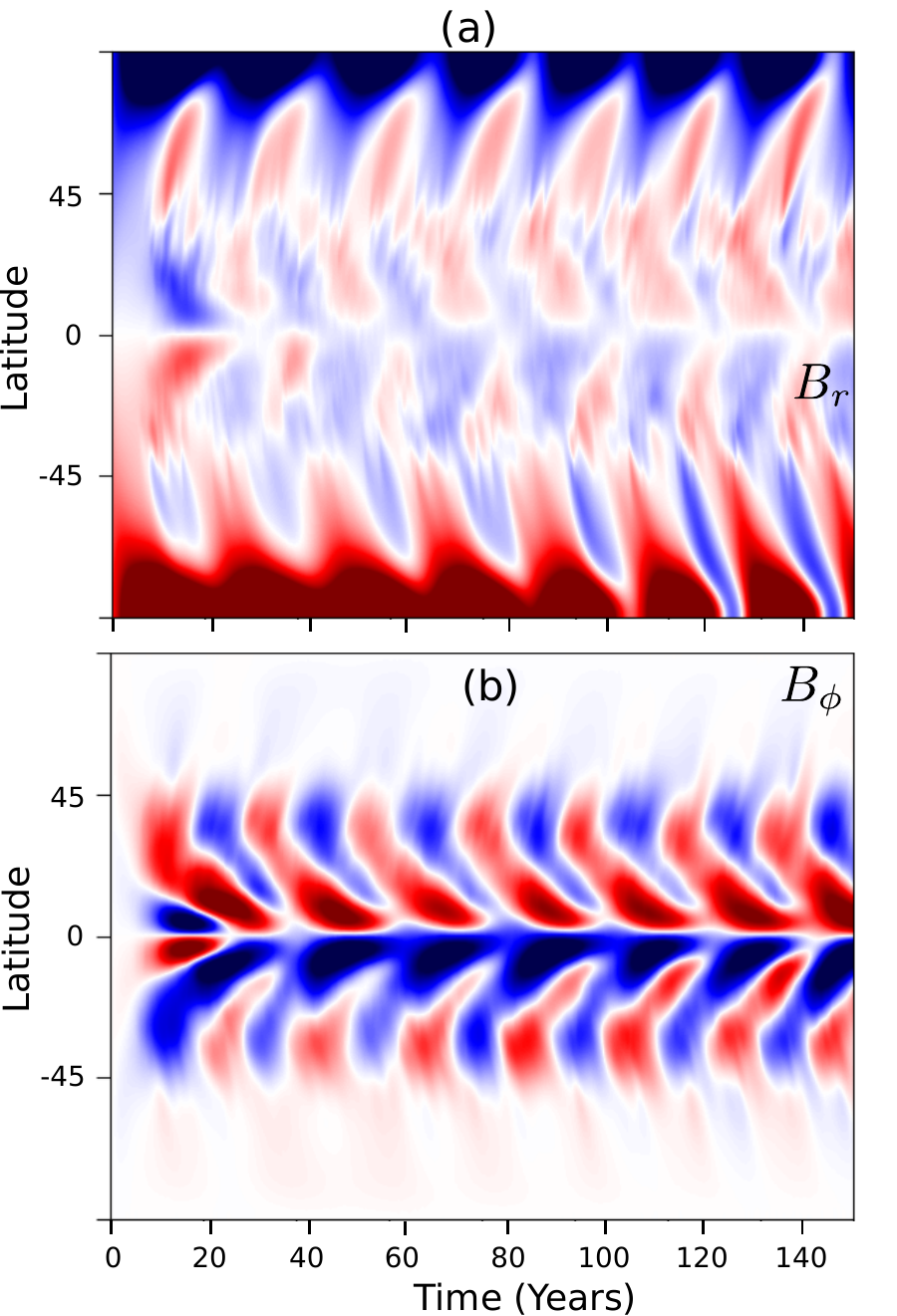}
\caption{For one BMR emerging every month: maps of the mean radial magnetic field ($B_r$) at the surface and the mean toroidal field ($B_\phi$) at the base of the convection zone.}
\label{fig:butterfly_bmr_1}
\end{figure}

The results presented so far were from the dynamo simulations in which 32 BMRs were allowed to emerge every month in the latitudinal region of $[-35, +35]$ degrees. Now we examine the effect of the number of emerging BMRs on the magnetic cycle by fixing all the other parameters ($V_{\theta p} = 12.5 \ \rm m \, \rm s^{-1}$ and $\eta_c = 2 \times 10^{10} \, \rm cm^2 \, \rm s^{-1}$). For this purpose, we perform simulations with 8 BMRs or 1 BMR emerging every month, compute the cycle amplitude in these cases, and then compare them with that of  the case with 32 BMRs emerging. In Table~\ref{table:table_var_bmr}, we present cycle duration, poloidal flux, and toroidal flux for different numbers of emerging BMRs. The table contains the averaged values of the aforementioned quantities for several magnetic cycles in the dynamo steady state. When we allow 8 BMRs to emerge every month, the duration of half a magnetic cycle remains almost the same as for 32 BMRs. However, the amplitude of the magnetic cycle is smaller. It is evident that when the total number of emerging BMRs is small, the magnitude of the resultant poloidal flux produced by the trailing spots is small, which leads to the generation of a smaller toroidal field for the subsequent cycle and hence the overall cycle amplitude is smaller. If we further decrease the number of emerging BMRs to one per month, the surface poloidal flux generated through the BL mechanism is not sufficient to reverse the polar field and hence we do not observe cyclic polar field reversals (see Fig.~\ref{fig:butterfly_bmr_1}a). We note however that in this case a toroidal field of polarity opposite to that of the polar cap gets produced in the convection zone. Therefore, we still get a strong toroidal field of a particular polarity at lower latitudes (see Fig.~\ref{fig:butterfly_bmr_1}b), which is due to the transport and winding of the strong polar field. However, not enough flux is taken up to the surface to reverse the polar field and the dynamo stays stationary, that is, it does not produce cycles. The dynamo fails when the number of BMRs emerging at the surface is further decreased. Therefore, it is important to have an adequate number of emerging BMRs in the system to sustain a proper magnetic cycle, which in our case is around eight BMRs every month.   

\begin{table}[htbp]
\caption{Duration and amplitude of the magnetic cycle with number of emerging BMRs every month.} 
\centering
\begin{tabular}{c c c c}
\hline 
\hline  
BMRs & $T_{1/2} $ & $\Phi(B_\phi)$ & $\Phi(B_r)$ \\ 
     & (Years) & $(10^{25} \rm Mx)$ & $(10^{25} \rm Mx)$ \\ [1ex]
\hline 
32 & $9.1$ & $1.9$ & $0.4$ \\
 
8 & $9.0$ & $1.5$ & $0.3$ \\
 
1 & $-$ & $-$ & $-$\\ [1ex]
\hline
\hline
\end{tabular}
\label{table:table_var_bmr} 
\end{table}

\subsection{Field-strength-dependent nonlinear buoyancy model}
\label{sec:tor_buoy}

In the earlier sections the velocity corresponding to the magnetic buoyancy was kept constant ($V_{b0} =94.5\; \rm m \, \rm s^{-1}$) for toroidal fields satisfying $B_{\phi}^l < B_\phi < B_{\phi}^h$. Here we implement a buoyancy velocity which varies with the amplitude of the toroidal field. Indeed, stronger flux tubes are supposed to be more buoyant and will thus rise faster, as confirmed by 3D numerical simulations of rising flux tubes in convective shells (e.g., \cite{Jouve:APJ2009}). The main purpose of employing this type of magnetic buoyancy is to observe if this nonlinearity introduces variability in the amplitude and the duration of the magnetic cycle, as previous mean-field calculations suggested~\citep{Jouve:AA2010b}. In the present case, $V_b \propto B_\phi^2$, that is, the buoyancy becomes more effective for a strong toroidal field and less effective for a weak toroidal field. For these simulations, we take $V_{\theta p} = 12.5 \, \rm m \, \rm s^{-1}$ and $\eta_c=2 \times 10^{10} \, \rm cm^2 \, \rm s^{-1}$. We compute parameters related to the magnetic cycle and present them in Table~\ref{table:table_var_vr}. In the first row, we show the cycle duration and amplitudes for a constant $V_b$ ($V_{b0}= 94.5 \; \rm m \, \rm s^{-1}$), whereas in the second and third rows, we present results with variable $V_b$. The cycle duration and amplitude change with the amplitude of the buoyancy velocity. When $V_{b0}$ ranges in $[7.8 - 94.5] \; \rm m \, \rm s^{-1}$ (rise time: 1 month - 1 Year), the cycle amplitude is slightly lower as compared to that for constant $V_{b0}$ at $94.5 \; \rm m \, \rm s^{-1}$. On the other hand, when $V_{b0}$ varies in $[12.3 - 153.0] \; \rm m \, \rm s^{-1}$ (rise times: 0.6 months - 7.5 months), we observe a slightly higher cycle amplitude compared to that for $V_{b0} = 94.5 \; \rm m \, \rm s^{-1}$. Our study suggests that when buoyancy is stronger (large $V_{b0}$), more toroidal flux gets buoyantly transported to the surface to produce BMRs which then generate a stronger poloidal field through the BL mechanism, which in turn produces a stronger toroidal field for the subsequent cycle. Therefore, a stronger buoyancy produces magnetic cycles with higher amplitudes. The results of these two simulations with variable magnetic buoyancy suggest that the cycle amplitude and the cycle strength do not change significantly for different cycles as compared to the case when the magnetic buoyancy is independent on the toroidal field strength. Therefore, our 3D kinematic model either seems to rule out field-strength-dependent buoyancy velocity as a strong source of amplitude modulation (or quenching), or our algorithm fails to adequately capture the full physics of nonlinearity in the process.  

\begin{table}[htbp]
\caption{Duration and amplitude of the magnetic cycle for a buoyancy velocity dependent on the toroidal field strength.} 
\centering
\begin{tabular}{c c c c}
\hline 
\hline  
$V_{b0}$ & $T_{1/2} $ & $\Phi(B_\phi)$ & $\Phi(B_r)$ \\ 
$(\rm m \, \rm s^{-1})$ & (Years) & $(10^{25} \rm Mx)$ & $(10^{25} \rm Mx)$ \\ [1ex]
\hline 
$94.5$ & $9.1$ & $1.9$ & $0.4$ \\
 
$7.8 - 94.5$ & $9.0$ & $1.8$ & $0.5$ \\
 
$12.3 - 153.0$ & $8.7$ & $2.1$ & $0.6$\\ [1ex]
\hline
\hline
\end{tabular}
\label{table:table_var_vr} 
\end{table}

\subsection{Effect of frequency of BMR emergence}

In the earlier simulations, a set of 32 new BMRs were made to emerge every month. Here we examine the impact of frequency of BMR emergence on the magnetic cycle at $V_{\theta p} = 12.5 \, \rm m \, \rm s^{-1}$ and $\eta_c=2 \times 10^{10} \, \rm cm^2 \, \rm s^{-1}$. In Table~\ref{table:table_var_freq_bmr}, we present cycle duration and amplitude for cases where 32 BMRs emerge every month, every 6 months, or every 12 months. We observe that the duration of half a magnetic cycle  anti-correlates with the frequency of emergence, that is, when there is a delay in the BMR emergence, the cycle duration becomes longer. If the BMRs emerge once in a 6- or 12-month period, we have less BMRs (or poloidal flux) at the surface and hence it takes longer for a sufficient amount of opposite magnetic flux to be advected towards the poles to reverse the polar field. On the other hand, the cycle amplitude is higher when the frequency of emergence is every 6 or 12 months. If the BMRs emerge less frequently, the toroidal field in the convection zone has a longer time available to amplify before getting buoyantly transported to the surface to produce BMRs with larger field strengths, which is the reason we observe a stronger cycle amplitude in these cases. We observe that the BMR field strength is higher in the cases when the frequency of emergence is once in every 6 or 12 months than in cases where the frequency of emergence is once a month. Also, the BMR field strength is higher when the frequency of emergence is once every 12 months as compared to when the frequency of emergence is once every 6 months. This result is somewhat counter intuitive but suggests that buoyant removal of toroidal flux has a limiting role on the cycle amplitude, as observed in earlier works \citep{Nandy2000}.

\begin{table}[H]
\caption{Duration and amplitude of the magnetic cycle with different frequencies of BMR emergence.} 
\centering
\begin{tabular}{c c c c}
\hline 
\hline  
Time between emergence of BMRs  & $T_{1/2} $ & $\Phi(B_\phi)$ & $\Phi(B_r)$ \\ 
 (months)     & (Years) & $(10^{25} \rm Mx)$ & $(10^{25} \rm Mx)$ \\ [1ex]
\hline 
$1$ & $9.1$ & $1.9$ & $0.4$ \\
 
$6$ & $9.4$ & $2.1$ & $0.5$ \\
 
$12$ & $10.2$ & $3.1$ & $0.7$\\ [1ex]
\hline
\hline
\end{tabular}
\label{table:table_var_freq_bmr} 
\end{table}

\subsection{A dynamo number characterizing the Babcock-Leighton dynamo} 

Here we propose an empirical dynamo number characterizing the Babcock-Leighton dynamo process ($D_{BL}$) that determines whether our model would produce self-sustained dynamo or not. Based on the parameter-space study presented in the previous sections, we observe that for a given rotation profile, fixed magnetic diffusivities at the top and bottom of the convection zone and fixed frequency of BMR emergence, the dynamo number would depend mainly on the convection zone diffusivity, strength of the meridional flow, and number of emerging BMRs at the surface. If the convection zone diffusivity is small, the dynamo will be more efficient, but if we go into the diffusion-dominated regime, the magnetic field will diffuse before getting significantly transported in the convection zone and hence the dynamo will likely fail. If the meridional circulation is very strong, it will also cause the decay of the magnetic field. The emerging BMRs play a crucial role in the magnetic cycle through their participation in the BL mechanism. If we do not have enough BMRs at the surface, then we do not observe cyclic field reversals. In addition, the tilt angle of a particular BMR decides how much flux the trailing spots contribute to the total surface poloidal flux. Considering all the aforementioned parameters, we define a Babcock-Leighton dynamo number
\begin{eqnarray}
D_{BL} = \frac{B_{eff} L_M}{\sqrt{\mu_0 \, \rho} \, \eta_c},
\end{eqnarray} 
which corresponds to a magnetic Reynolds number calculated with an Alfv\'en velocity and a well-chosen characteristic length scale. Here we define $B_{eff} = \frac{\Phi_{eff}}{A_{Polar Cap}}$ ($\Phi_{eff}$ is the magnitude of the total magnetic flux at the polar cap or the effective flux of all the trailing spots that get transported towards the poles to cancel the polar flux, and $A_{Polar Cap}$ is the surface area covered by the region $[0^\circ - 20^\circ]$ colatitudes) and $\eta_c$ is the convection zone magnetic diffusivity. We also define an effective flux transport length-scale $L_M = \sqrt{\eta_s \, t_M}$, where $t_M = \frac{\pi R_{\odot}}{2 V_{\theta p}}$ is the time it takes for a fluid particle to be advected from the equator to the pole by the meridional flow at the surface. This length scale is simply the convection-zone length scale when the diffusion and advection timescales are equal, but is shorter in the advection-dominated regime compared to the diffusion-dominated regime (i.e., when the advection timescale is shorter than the diffusion timescale). Finally, we consider here that $\sqrt{\mu_0 \, \rho} =1$.
Hence,
\begin{eqnarray}
D_{BL} = \frac{\Phi_{eff}}{A_{Polar Cap}} \frac{\sqrt{\eta_s \, t_M}}{\eta_c}.
\end{eqnarray}
 
In our simulations, the length of half a solar cycle is
\begin{eqnarray}
\tau_{1/2} \approx 50.0 \; V_{\theta p}^{-0.65}.
\end{eqnarray}

Therefore, the total number of BMRs at the surface in $\tau_{1/4}$ (time available for polar field reversal)
\begin{eqnarray}
N_{BMR}(\tau_{1/4}) & \approx & 25.0  \; V_{\theta p}^{-0.65} N_{BMR} Fr_{BMR}  \nonumber \\
& = & 300 \; V_{\theta p}^{-0.65} N_{BMR},
\end{eqnarray}
where $N_{BMR}$ is the number of BMRs emerging at one instance and $Fr_{BMR}$ is the frequency of emerging $N_{BMR}$ BMRs (once every month or 12 times per year). 

As the trailing spots are at slightly higher latitudes, we assume that only the magnetic flux of the higher-latitude section  of a trailing spot gets transported towards the poles and that the rest gets annihilated due to the cross-equatorial cancellation and by the local opposite-polarity magnetic flux through diffusion. We define the magnetic flux of the section of a trailing spot which is at higher latitudes as the effective flux of one BMR. To make things simple, we consider the average tilt angle of BMRs to be $10$ degrees. The average effective flux of one BMR comes out to be $4.29 \times 10^{21}$ Mx and therefore the dynamo number becomes 
\begin{eqnarray}
D_{BL} = 2.33 \times 10^{14} \; N_{BMR} \frac{V_{\theta p}^{-1.15}}{\eta_c},
\end{eqnarray}
where $V_{\theta p}$ is in the units of $\rm cm \, \rm s^{-1}$ and $\eta_c$ in $\rm cm^2 \, \rm s^{-1}$. For example, if $N_{BMR} =32$, $V_{\theta p} =12.5 \; \rm m \, \rm s^{-1}$ (or $1250 \; \rm cm \, \rm s^{-1}$), and $\eta_c =2.0 \times 10^{10} \; \rm cm^2 \,\rm s^{-1}$, we obtain $D_{BL} \approx 103$. Below a critical value of $D_{BL}$ the system would not be able to sustain the dynamo and hence the magnetic cycle through flux-transport in the convection zone. In our model the estimated value of $D_{BL}^c$ is approximately $5$. It is apparent that dynamo action may not be sustained for a very strong meridional circulation or for a highly diffusive convection zone or for a very small number of emerging BMRs at the surface. In our model, dynamo works for (a) $N_{BMR} =8$, $V_{\theta p} =12.5 \; \rm m \, \rm s^{-1}$, $\eta_c =2.0 \times 10^{10} \; \rm cm^2 \, \rm s^{-1}$, and ($D_{BL} \approx 25.6$), and (b) $N_{BMR} =32$, $V_{\theta p} =40 \; \rm m \, \rm s^{-1}$, $\eta_c =2.0 \times 10^{10} \; \rm cm^2 \, \rm s^{-1}$, and ($D_{BL} \approx 26.9$). On the other hand, dynamo fails for (a) $N_{BMR} =1$, $V_{\theta p} =12.5 \; \rm m \, \rm s^{-1}$, and $\eta_c =2.0 \times 10^{10} \; \rm cm^2 \, \rm s^{-1}$, ($D_{BL} \approx 3.2$), and (b) $N_{BMR} =32$, $V_{\theta p} =80 \; \rm m \, \rm s^{-1}$, and $\eta_c = 10^{11} \; \rm cm^2 \, \rm s^{-1}$ ($D_{BL} \approx 2.4$). We thus confirm that this evaluation of the dynamo number in our particular model is well adapted to anticipate the growth of magnetic energy when the meridional flow speed, convection-zone diffusivity, and number of BMRs are given. We note that for the calculation of the dynamo number, we considered a uniform tilt angle for all the bipolar spots. The effect of tilt angle is included in the calculation of the average effective flux of one BMR where for simplicity we considered the average tilt angle of all the BMRs to be $10$ degrees. If we were to consider a larger (smaller) tilt angle, we would have a higher (lower) effective flux and hence a larger (smaller) dynamo number. It is important to note that this dynamo number determination should be taken as specific to this model and has scope for reformulation for models with diverse physics.
 
\section{Discussion and conclusions}
\label{sec:conclusion}

In this paper, we present a global 3D kinematic solar dynamo model to explore flux transport processes that sustain the Babcock-Leighton dynamo. We have used a solar-like differential rotation and meridional circulation as the prescribed velocity field for the kinematic dynamo simulations and a parametrized turbulent diffusivity to characterize the solar interior.  We have implemented a 3D magnetic buoyancy algorithm in the convection zone which transports strong toroidal magnetic fields from the base of the convection zone to the surface to produce tilted BMRs. The erupted BMRs subsequently generate a poloidal field at near-surface layers which reverse the old cycle poloidal field. This newly generated poloidal field is then subducted to deeper layers of the convection zone, where differential rotation then stretches it to generate the toroidal field of the next sunspot cycle. Magnetic flux-transport plays an important role in these different processes resulting in long-term cyclic polarity reversals driven by the active participation of tilted, bipolar sunspot pairs. We note that our relatively more realistic methodology of modeling active-region emergence through an effectively 3D helical flow -- consequent BMR formation, and self-sustained cyclic reversal over multiple cycles -- provides critical advances over 2D kinematic dynamo models and complements 3D global MHD models of the solar cycle. 

There are various flux-transport parameters involved in the solar dynamo which affect the nature of the magnetic cycle, especially their duration and amplitudes. To understand these aspects of the solar dynamo, we varied the strength of the meridional circulation, convection zone diffusivity, and parameters related to BMR emergence. The major findings of our parameter-space studies are as follows.

\begin{enumerate}[ {}1{.} ]
\item The duration of the magnetic cycle is highly sensitive to the strength of the meridional circulation where cycle duration is shorter for a stronger meridional flow. However, for a diffusion-dominated regime (large convection zone diffusivity $\eta_c$ and small peak meridional speed $V_{\theta p}$), the cycle duration depends less on the meridional circulation, because in this case the magnetic diffusion also plays an important role in the flux-transport process.
\item For an advection-dominated situation (low convection zone diffusivity $\eta_c$ and high peak meridional speed $V_{\theta p}$), we observe a lower cycle amplitude as the circulation speed is increased. This is because a stronger meridional flow drags the poloidal component quickly through the generating layer of the toroidal component, inducing a weaker toroidal field.
\item In the diffusion-dominated regime, however, the cycle amplitude is higher for stronger meridional flow. This is because for a higher circulation speed, the poloidal field  being subducted in to the convection zone suffers less diffusive decay, which results in a larger source for the toroidal field in the rotational shear layers.
\item In the advection-dominated regime, the poloidal field of cycles $n-2$, $n-1,$ and $n$ combine to produce the toroidal field of cycle $n+1$, which suggests that the memories of the previous few cycles propagate to the subsequent cycle. We do not observe this kind of memory propagation in the diffusion-dominated case where the memory is limited to only one cycle. This feature is in agreement with previous 2D calculations and is important for forecasting future solar activity.
\item The cycle duration does not change with the number of emerging BMRs, but the cycle amplitude depends on it. If the number of emerging BMRs is less, then the amplitudes of the magnetic cycles are smaller. The dynamo fails when the number of emerging BMRs is too small as this results in a very weak seed for the poloidal field in the context of the BL mechanism. The resultant weak poloidal field is insufficient to reverse the polar field and sustain new cycles. 
\item The amplitude of the buoyancy velocity affects the cycle amplitude. Faster buoyancy velocity in the convection zone produces  magnetic cycles with slightly higher amplitudes. This is because toroidal flux is transported to the surface at a faster rate. The resultant BMRs are stronger, less diffused, and more coherent, producing stronger poloidal fields through the BL process. 
\item If the BMRs emerge less frequently, the cycle duration as well as the cycle amplitude increases. It takes longer for a sufficient number of BMRs to emerge and hence poloidal flux to appear at the surface, which would then participate in polar-field reversals. This is the reason why we observe a longer cycle duration. Also, if the frequency of emergence is small, the toroidal field near the base of the convection zone gets amplified for a longer time before being buoyantly transported to the surface and therefore we observe a stronger cycle amplitude in these cases. This magnetic buoyancy may play a role as an amplitude-limiting mechanism under certain conditions.  
\item We have defined a (empirical) dynamo number for our Babcock-Leighton model which depends on the strength of the meridional circulation, convection zone diffusivity, and number of emerging BMRs. There is a critical dynamo number below which the BL dynamo action becomes unsustainable. Our extensive numerical simulations show that dynamo action cannot be excited for too strong a meridional circulation, very large convection zone diffusivity, very few emerging BMRs, or a combination of these.  The determined BL dynamo number enables us to anticipate whether or not dynamo action will indeed be sustainable for various governing parameters in this model.          
\end{enumerate}

In conclusion, the 3D kinematic Babcock-Leighton dynamo model presented here demonstrates many aspects of the solar magnetism and in particular the role of flux-transport processes in the sustenance of the sunspot cycle.

A comparative assessment of this study vis-a-vis understanding gained from previous studies may be illuminative. On the one hand, some of the conclusions drawn from this 3D model are in broad agreement with results from 2D models in the context of advection- versus turbulent-diffusion-dominated convection zones with profound implications for solar-cycle memory and predictions \citep{Yeates:APJ2008}. This is reassuring. On the other hand, some of the model elements are important, additional insights from this
model: for example the effective 3D helical flows utilized in modelling magnetic buoyancy and tilted bipolar sunspot pair formation, the impact of buoyant-flux-transfer rate (i.e., amplitude of effective buoyancy velocity), and the frequency of emergence in determining the sunspot-cycle properties. The ideal parameter defining the efficiency of diverse solar dynamo models is the dynamo number. However, in the context of Babcock-Leighton dynamos, it is not immediately apparent what this should be because the source is modeled differently from in traditional mean-field dynamos. We have made a first attempt at empirically defining a dynamo number specific to the flux transport Babcock-Leighton solar dynamo model which reasonably captures our model dynamics.  We also note that as compared to \cite{Yeates:MNRAS2013} who simulated only one cycle, we successfully excite self-sustained multi-cycle dynamo action with a field-strength-dependent buoyancy algorithm modeled through helical upflows.

We note a few shortcomings of our model. For example, we observed a significant overlap between two consecutive cycles which prohibits us from achieving proper solar minimum-like conditions in between cycles. In our model, the surface magnetic field is high compared to the observations. These are typical, known issues in 2D flux-transport models that seem to carry over to 3D global models, implying that their solution necessitates other considerations. 
  
Our detailed parameter-space studies reveal the characteristics of magnetic cycles in the advection- and diffusion-dominated regimes, which is crucial to understanding variations in solar-cycle amplitude, solar-cycle duration and memory propagation from one cycle to another. The latter in particular is deemed crucial for solar-cycle predictions and for determining the nature of poloidal field input for driving predictive dynamo models.  We note that the surface radial magnetic field produced in our 3D flux-transport dynamo model can be directly used to study the evolution of coronal structures (see e.g., \cite{Nandy2018}) and the solar wind (see e.g.,  \cite{Kumar:Frontiers2018}) over solar-cycle timescales and is amenable to observational corrections. This 3D global dynamo model is also amenable to data-assimilation-based prediction of future solar activity through direct forcing at the solar surface with observed flows and observed poloidal field. This model can also be adapted to study stellar magnetic activity and magnetized starspot formation in other stars with different internal structures and plasma flow profiles. The versatile nature of this model may be useful for addressing these diverse practical applications in the future. 

\begin{acknowledgements}
This work was supported by the Indo-French research grant 5004-1 from CEFIPRA/IFCPAR. Numerical simulations were performed at CINES, the supercomputing facility in Montpellier, France, and at the CALMIP supercomputing facility at the Universit{\'e} de Toulouse (Paul Sabatier), France. The Center of Excellence in Space Sciences India is funded by the Ministry of Human Resource Development of the Government of India. We thank Prantika Bhowmik for useful discussions and suggestions. L.J. acknowledges support from the Institut Universitaire de France (IUF).
\end{acknowledgements}

\bibliographystyle{aa} 
\bibliography{solar_dyn} 


\end{document}